\documentclass[twocolumn,english,aps,pra,manuscript,reprint,superscriptaddress,showpacs,showkeys,longbibliography]{revtex4-1}
\usepackage[T1]{fontenc}
\usepackage[latin9]{inputenc}
\usepackage{color}
\usepackage{babel}
\usepackage{amsmath}
\usepackage{amssymb}
\usepackage{graphicx}
\usepackage[colorlinks=true,
    linkcolor=blue,
    citecolor=blue,
    urlcolor =blue]{hyperref}
\makeatletter
\usepackage{amsfonts}
\usepackage{babel}
\usepackage{braket}
\makeatother

\newcommand{\fref}[1]{Fig.\,\ref{#1}}
\newcommand{\Fref}[1]{Figure \ref{#1}}
\newcommand{\eref}[1]{Eq.\,\eqref{#1}}

\newcommand{\lr}[1]{\left(#1\right)}
\newcommand{\dg}{^{\dag}}

\begin{document}

\title{Quantum optics of chiral spin networks}
\author{Hannes Pichler}
\email{hannes.pichler@uibk.ac.at}
\affiliation{Institute for Quantum Optics and Quantum Information of the Austrian
Academy of Sciences, 6020 Innsbruck, Austria}
\affiliation{Institute for Theoretical Physics, University of Innsbruck, 6020 Innsbruck, Austria}

\author{Tom\'as Ramos}
\affiliation{Institute for Quantum Optics and Quantum Information of the Austrian
Academy of Sciences, 6020 Innsbruck, Austria}
\affiliation{Institute for Theoretical Physics, University of Innsbruck, 6020 Innsbruck, Austria}

\author{Andrew J. Daley}
\affiliation{Department of Physics and SUPA, University of Strathclyde, Glasgow
G4 0NG, UK}

\author{Peter Zoller}
\affiliation{Institute for Quantum Optics and Quantum Information of the Austrian
Academy of Sciences, 6020 Innsbruck, Austria}
\affiliation{Institute for Theoretical Physics, University of Innsbruck, 6020 Innsbruck, Austria}
\affiliation{Max-Planck-Institut f\"ur Quantenoptik, 85748 Garching, Germany}

\begin{abstract}
We study the driven-dissipative dynamics of a network of spin-1/2 systems coupled to one or more chiral 1D bosonic waveguides within the framework of a Markovian master equation. We determine how the interplay between a coherent drive and collective decay processes can lead to the formation of pure multipartite entangled steady states. 
The key ingredient for the emergence of these many-body dark states is an asymmetric coupling of the spins to left and right propagating guided modes. Such systems are motivated by experimental possibilities with internal states of atoms coupled to optical fibers, or motional states of trapped atoms coupled to a spin-orbit coupled Bose-Einstein condensate. We discuss the characterization of the emerging multipartite entanglement in this system in terms of the Fisher information. \end{abstract}
\maketitle

\section{Introduction}

\begin{figure}[t!]
\includegraphics[width=0.45\textwidth]{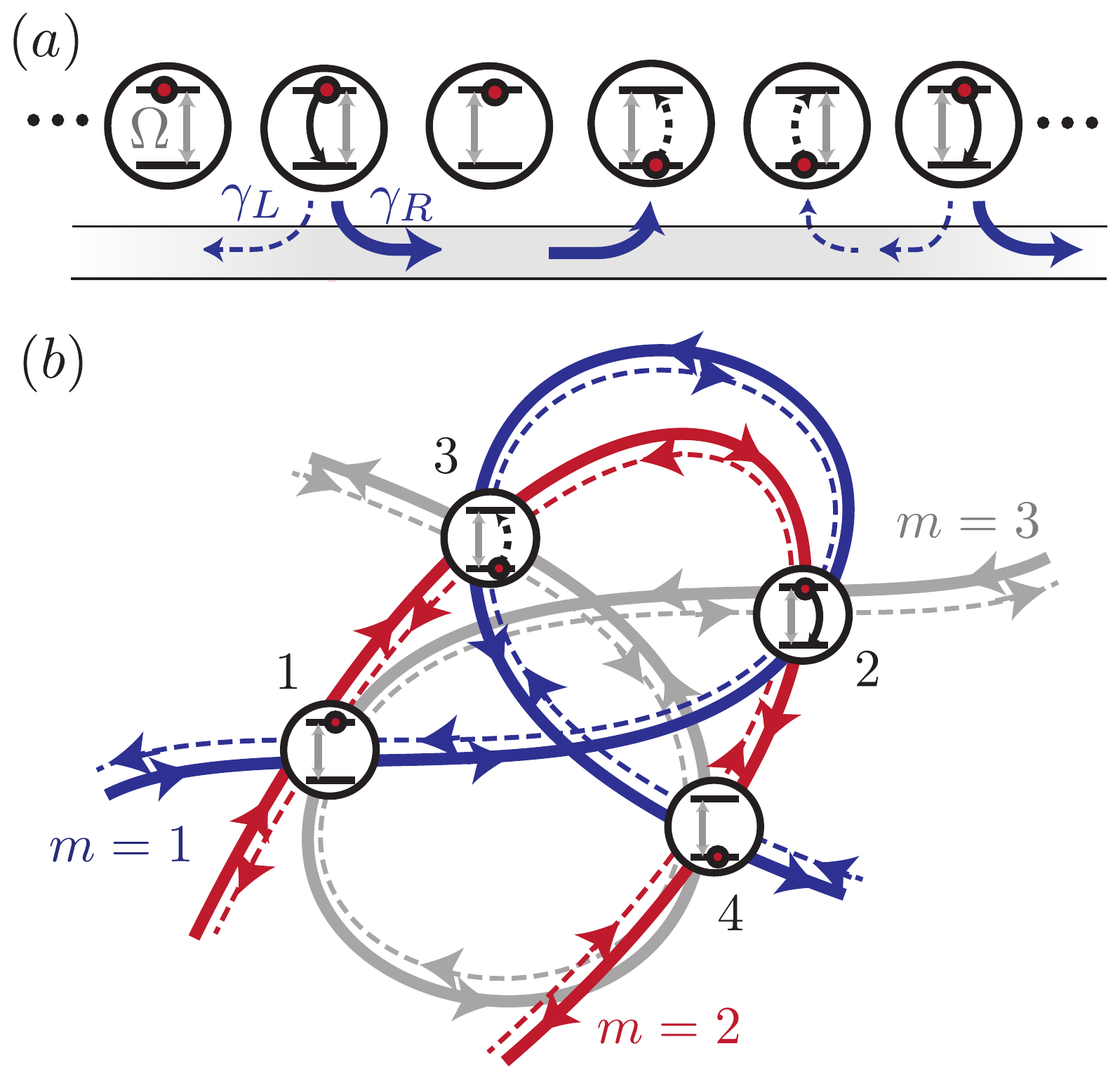}
\caption{(Color online) Spin networks with chiral coupling to 1D bosonic reservoirs. (a) Driven spins can emit photons to the left and right propagating reservoir modes, where the chirality of the system-reservoir interaction is reflected in the asymmetry of the corresponding decay rates $\gamma_L\neq\gamma_R$. (b) Spin network coupled via three different chiral waveguides $m=1,2,3$. Waveguide $m=1$ couples the spins in the order $(1,2,3,4)$, whereas $m=2$ couples them in order $(1,3,2,4)$ and $m=3$ in order $(2,1,4,3)$. Note that only waveguides without closed loops are considered in this work.}
\label{fig:model}
\end{figure}

The ability to engineer the system-bath coupling in quantum optical systems allows for novel scenarios of dissipatively preparing quantum many-body states of matter \cite{Muller:2012wh}. This is of interest both as a nonequilibrium condensed matter physics problem \cite{Diehl:2008aa,Knap:2013el,Rao:2013de,Carr:2013hc,Honing:2013fx,Scelle:2013in,Chen:2014it} and in the context of quantum information \cite{Krauter:2011fj,Weimer:2010ez,Verstraete:2009kc,Barreiro:2011jq,Lin:2013cc,Shankar:2013do,vanLoo:2013df,Kastoryano:2013if,Kastoryano:2011hr,Vollbrecht:2011kv}. In the present work we will study open system quantum dynamics of {\em chiral spin networks} from a quantum optical perspective. The nodes of these networks are represented by spin-1/2 systems, whereas the quantum channels connecting them are 1D waveguides carrying bosonic excitations [cf.~Fig.\,\ref{fig:model}(a) and \ref{fig:model}(b)]. In addition, these waveguides provide the input and output channels of our quantum network, allowing for driving and continuous monitoring of the spin dynamics. In a quantum optical setting, such a network can be identified by two-level atoms coupled to optical fibers \cite{Reitz:2013bs,Yalla:2014eb} or photonic structures \cite{Thompson:2013hx,Goban:2013wp}. As discussed in previous studies \cite{Chang:2012co,GonzalezTudela:2013hn,Schneider:2002bh}, the 1D character of the quantum reservoir manifests itself in unique features including long-range dipole-dipole interactions mediated by the bath and collective decay of the two-level systems as super- and subradiant decay.

The crucial aspect underlying our study below is the assumption of a {\em chiral} character of the waveguides representing the photonic channels. By chirality we mean that the symmetry of emission of photons from the two-level atoms into the right and left propagating modes of the 1D waveguides is broken. This allows the formation of novel nonequilibrium quantum phases as steady states of the open system dynamics in chiral quantum spin networks. This includes the driven-dissipative evolution as ``cooling'' to {\em pure  states} of {\em entangled spin clusters}, which play the role of quantum {\em many-particle dark states}, i.e. spin clusters decoupled from the bath. While in Ref.~\cite{Stannigel:2012jk} the formation of entangled spin clusters for the (idealized) purely unidirectional waveguide has been discussed, we have recently presented results that this formation of pure entangled spin clusters is, in fact, the generic case for chiral spin networks under fairly general conditions \cite{Ramos:2014ut}. It is the purpose of the present paper to present an in depth study of this quantum dynamics and pure entangled spin cluster formation in chiral spin networks including imperfections, and the characterization of the resulting multipartite entangled states in experiments (e.g., via the Fisher information \cite{Hyllus:2012ha,Strobel:2014eg}). We emphasize that our results are derived within the standard quantum optical master equation (ME) treatment, where the effective spin dynamics is obtained by eliminating the reservoir in a Born-Markov approximation (in contrast to non-Markovian treatments discussed, for instance, in Refs.~\cite{Bre2009aa,Huelga:2012di,Zha2012aa,Douglas:2013tb,Bylicka:2014hh}).

The present work is motivated by recent experiments and proposals for the realization of chiral spin networks with quantum optical systems. This includes the remarkable recent experimental demonstration of directional spontaneous emission of single atoms and scattering from nanoparticles into a 1D photonic nanofiber in Refs.\,\cite{Petersen:2014tc,Mitsch:2014tr}, and similar experiments and proposals with quantum dots coupled to photonic nanostructures in Refs.\,\cite{Sollner:2014wb,Young:2014ty}. We also note that related experiments reporting directional emission in 2D setups have been performed with photons \cite{Neugebauer:2014iy} and surface plasmons \cite{RodriguezFortuno:2013ew,Lin:2013do}. Moreover we remark that topological photonic devices provide chiral edge modes for light propagation \cite{Hafezi:2013jg,Rechtsman:2013cz}, with possible applications in this context. In contrast to these {\em photonic} setups, we have shown in Ref. \cite{Ramos:2014ut} that chiral waveguides for {\em phonons} (or Bogoliubov excitations) can be realized using a 1D spin-orbit coupled Bose-Einstein condensate (SOC BEC) \cite{Lin:2011hn,Galitski:2013dh}. A faithful realization of the corresponding chiral spin network was proposed by coupling atoms in optical lattices, representing spins with vibrational levels, to the SOC BEC via collisional interactions.

The paper is organized as follows. In Sec.\,\ref{modeloverview} we present the quantum optical model describing the chiral spin networks and provide a qualitative summary of the various multipartite entangled pure states, which are formed as steady states of their driven-dissipative dynamics. In Sec.\,\ref{steadystates} we illustrate this for networks of two and four spins and identify sufficient conditions for the existence of pure steady states as  "dark states" of the quantum master equation. In Sec.\,\ref{Nspins} we extend these concepts to networks with an arbitrary number of spins. There, we will also analyze the purification dynamics and comment on the role of imperfections. In Sec.\,\ref{FisherSec}, we discuss the possibility to witness the steady-state multipartite entanglement via a measurement of the Fisher information. In Sec.\,\ref{outlook} we conclude with an outlook.

\section{Model and Overview}\label{modeloverview}

The key feature of the chiral spin networks considered here is the accessibility of pure multipartite entangled states that arise as the steady state of their driven-dissipative dynamics. In this section we give an overview of this primary result, beginning with an introduction to the underlying physical model of the chiral spin network itself. We then discuss the master equations that describe the corresponding open system dynamics and illustrate the entanglement properties of their pure steady states in different parameter regimes.

\subsection{The chiral spin network}

The system we consider consists of a collection of $N$ two-level systems (TLSs) or \emph{spins}, as depicted in \fref{fig:model}(a). For each spin $j$, we will denote the two states by $\ket{g}_j$ and $\ket{e}_j$, and the corresponding transition frequency between the two states by $\omega_{j}$. These spins are driven by a classical coherent field at a single frequency $\nu$, defining a detuning pattern $\delta_j\equiv\nu-\omega_j$. We denote the corresponding Rabi frequencies by $\Omega_j$. In a rotating frame with the driving frequency $\nu$ and after applying the rotating wave approximation (RWA), provided $|\Omega_j|,|\delta_j|\ll \omega_{j}$, the Hamiltonian for the spin chain reads
\begin{align}
H_{\rm sys}=\hbar\sum_{j=1}^N\big(\!-\!\delta_{j} \sigma_j^{\dag}\sigma_j+\Omega_j\sigma_j+\Omega_j^{\ast}\sigma_j^\dag\ \big),\label{Hsystem}
\end{align}
where we have used the notation $\sigma_j\equiv\ket{g}_j\!\bra{e}$. The spins are coupled to a 1D waveguide, whose Hamiltonian is given by 
\begin{align}
H_{\rm res}=\sum_{\lambda=L,R}\int\! d\omega\ \hbar\omega \,b_{\lambda}^{\dag}(\omega)b_{\lambda}(\omega),\label{Hreservoir}
\end{align}
where the $b_\lambda(\omega)$ are bosonic annihilation operators for the right ($\lambda\!=\!R$) and left ($\lambda\!=\!L$) moving bath modes of frequency $\omega$; i.e.,~$[b_\lambda(\omega),b_{\lambda'}^{\dag}(\omega')]=\delta_{\lambda,\lambda'}\delta(\omega-\omega')$. We note that in writing \eref{Hreservoir} we implicitly assumed a linear dispersion relation for the degrees of freedom of the reservoir. 

We are interested in a \emph{chiral coupling} of the spins to these reservoir modes. By this we refer to an asymmetry in the coupling to the left and right propagating modes of the waveguide. Such a chiral system-reservoir interaction can be modeled by the linear RWA Hamiltonian
\begin{align}
H_{\rm int}\!=\!i\hbar\!\!\!\sum_{\lambda=L,R}\sum_{j}\!\int\!\! d\omega\sqrt{\frac{\gamma_\lambda}{2\pi}}b_\lambda^{\dag}(\omega)\sigma_je^{-i(\nu t+\omega x_j/v_\lambda)}+\rm h.c.,\label{chiralInt}
\end{align}
where $\gamma_L$ and $\gamma_R$ are the decay rates into the left ($v_L\!<\!0$) and right ($v_R\!>\!0$) moving reservoir modes, respectively, with $v_\lambda$ denoting the corresponding group velocities. In addition, we denote the position of spin $j$ along the waveguide by $x_j$. We stress the fact that the chirality of the system-reservoir coupling is reflected by $\gamma_L\neq\gamma_R$.

\subsection{Chiral dissipative dynamics and overview of parameter regimes}\label{mastereqSec}
Treating the chiral waveguide as a long reservoir exhibiting Markovian dynamics, we can now derive a master equation describing the dissipative dynamics of the spin degrees of freedom. If we make the standard quantum-optical Born-Markov approximation and neglect retardation effects (which is valid provided $|\Omega_j|,\gamma_j,|\delta_j|\!\ll\!|v_\lambda|/|x_j-x_l|,\omega_j$), we obtain a master equation for the evolution of the system density operator $\rho(t)$, as detailed in Appendix~\ref{DerivationME}. Using the notation $\mathcal{D}[A]\rho\equiv A\rho A\dg\!-\!\{A\dg A,\rho\}/2$, the chiral master equation in explicit Lindblad form reads
\begin{align}
\dot{\rho}=-\frac{i}{{\hbar}}[H_{\rm sys}\!+\!H_L\!+\!H_R,\rho]\!+\!\gamma_L\mathcal{D}[c_L]\rho\!+\!\gamma_R\mathcal{D}[c_R]\rho,\label{chiralME}
\end{align}
where left and right moving reservoir modes give independent contributions. Their coherent parts 
\begin{align}
H_{L}&\equiv-\frac{i\hbar\gamma_{L}}{2}\sum_{j<l}\lr{e^{ik|x_j-x_l|}\sigma_j\dg \sigma_l-\rm h.c.},\label{Hleft}\\
H_{R}&\equiv-\frac{i\hbar\gamma_{R}}{2}\sum_{j>l}\lr{e^{ik|x_j-x_l|}\sigma_j\dg \sigma_l-\rm h.c.},\label{Hright}
\end{align}
describe long-range spin interactions mediated by the left and right moving reservoir modes, respectively. Due to the 1D character of the bath these interactions are of infinite range. However, the positions of the spins $x_j$ enter due to their \emph{ordering} along each propagation direction. Without loss of generality we label the spins such that $x_j>x_l$ for $j\!>\!l$. The second relevant quantity thereby is their distance as compared to the wavevector $k$ of the resonant reservoir modes (cf.~Appendix~\ref{DerivationME}). The dissipative terms with collective jump operators $c_L\!\equiv\!\sum_j e^{ikx_j}\sigma_j$ and $c_R\!\equiv\!\sum_j e^{-ikx_j}\sigma_j$ describe collective spin decay into left and right moving excitations that leave the waveguide at the two different output ports [cf.~\fref{fig:model}(a)]. Therefore, in contrast to the coherent part, the dissipative part does not depend on the ordering of the spins along the waveguide. 

In the rest of this subsection we discuss the two limiting cases corresponding to a bi-directional (nonchiral) situation $\gamma_L=\gamma_R$, and a purely cascaded one where $\gamma_L=0$. We then introduce a more general situation considering multiple chiral waveguides. 

\subsubsection{Bidirectional master equation}\label{dickemodel}

We note that the familiar Dicke model \cite{Gross:1982js} in one dimension is obtained from the chiral master equation (\ref{chiralME}) in the limit of a perfect bidirectional reservoir; i.e., when the symmetry between left and right moving excitations is not broken, $\gamma_L\!=\!\gamma_R\!\equiv\!\gamma$. In this case, $H_L+H_{R}$ conspires to form the well-known infinite-range dipole-dipole Hamiltonian, whereas the Lindblad terms form the familiar super- and sub-radiant collective decay \cite{{Lehmberg:1970jj}} 
\begin{align}\label{bidirectionalME}
\dot\rho&=-\frac{i}{\hbar}\Big[H_{\rm sys}+\hbar\gamma\sum_{j,l} \sin(k|x_j-x_l|) \sigma_j^{\dag}\sigma_l,\rho\Big]\nonumber\\&+2\gamma\sum_{j,l} \cos(k|x_j-x_l|) \lr{\sigma_l\rho \sigma_j^{\dag}-\frac{1}{2}\{\sigma_j\dg \sigma_l,\rho\}}.
\end{align}
Both coherent and dissipative parts depend on the distance between spins only up to a multiple of the wavelength. Therefore, in contrast to the chiral situation, the order of the spins does not matter. 

Remarkably, when placing the spins at distances commensurate with the reservoir wavelength such that $k|x_j-x_l|=2\pi n$ ($n$ integer), the dipole-dipole interactions vanish and the collective jump operators to left and right moving excitation modes coincide $c_L=c_R=\sum_j \sigma_j\equiv c$. When driving all spins homogeneously $\Omega_j=\Omega$ and on-resonance $\delta_j=0$, this reduces to a totally symmetric Dicke model \cite{GonzalezTudela:2013hn,Walls:1978hh}
\begin{align}
\dot\rho&=-i[\Omega(c+c^\dag),\rho]+2\gamma{\cal D}[c]\rho. \label{purelydissipativeDicke}
\end{align}
This model is symmetric under exchange of all the spins, giving rise to multiple steady states corresponding to decoupled subspaces in different symmetry sectors. On each of these subspaces, the system of $N$ spin-1/2s reduces to a single collective spin-$J$, where $J=0,1,\dots,N/2$ (for even $N$) is determined by the initial condition. Interestingly, this model predicts a non-equilibrium phase transition, e.g. in the $J=N/2$ manifold, at a critical driving strength $\Omega_c\equiv N\gamma/4$ \cite{GonzalezTudela:2013hn,Walls:1978hh}.

\subsubsection{Cascaded master equation}

The other limiting case of a chiral waveguide is a purely unidirectional reservoir, where the spin chain couples only to modes propagating in one direction (e.g. $\gamma_L=0$). One refers to such a system as \emph{cascaded}, since the output of each spin can only drive other spins located on its right, without back-action. The corresponding cascaded master equation was extensively studied in Refs.~\cite{{Carmichael:1993el},{Gardiner:1993cy},Gardiner:1994jz,{Stannigel:2012jk},{Cirac:1997is},{Stannigel:2010eu}} and it is simply given by setting $\gamma_L=0$ and thus $H_L=0$ in Eq.\,(\ref{chiralME}). To gain more insight into the dynamical structure of such a unidirectional channel, we rewrite \eref{chiralME} for this specific case as
\begin{align}\label{Cascaded_ME}
\dot\rho&=-\frac{i}{\hbar}[H_{\rm sys},\rho]-\frac{i}{\hbar}(H_{\rm eff}\rho-\rho H_{\rm eff}\dg)+\gamma_R c\rho c\dg,
\end{align}
where the non-Hermitian effective Hamiltonian reads
\begin{align}
H_{\rm eff}=-\frac{i\hbar\gamma_R}{2}\sum_j \sigma_j\dg \sigma_j-i\hbar\gamma_R\sum_{j>l}\sigma_j\dg \sigma_l.\label{nonHermitian}
\end{align}
To connect to the standard literature we have (without loss of generality) absorbed phases by $\sigma_j\rightarrow \sigma_je^{ikx_j}$ and $\Omega_j\rightarrow \Omega_je^{-ikx_j}$. The positions of the spins then enter solely via their spatial ordering. Note that such a simplification is only possible in the strict cascaded case, since there is only one collective jump operator, by construction. Between the corresponding quantum jumps \cite{QuantumNoise}, the system evolves with the non Hermitian Hamiltonian in Eq.~(\ref{nonHermitian}). It induces \emph{unidirectional} interactions between spins, where an excitation of spin $l$ can be transferred \emph{only} to spins $j$ located on its right ($j>l$). The inverse process is not possible. In contrast to conventional spin interactions \cite{Auerbach:1994jf}, these unidirectional interactions are thus fundamentally non-Hermitian, and can not be obtained in a closed system.

\subsubsection{Multiple-waveguide chiral master equation}

In a more general context one can consider spins coupled not only by one, but by several chiral waveguides as depicted in Fig.~\ref{fig:model}(b). These additional waveguides, labeled by  $m=1,...,M$, are arranged such that the order of the spins along each of them differs. We are interested in the situations where each of these waveguides couples to each spin at most once, excluding, for example, loops. Since the different waveguides are independent, it is straightforward to generalize the chiral ME from a single- to a multiple-waveguide network, where each waveguide gives an additive contribution analogous to Eq.~\eqref{chiralME}. Denoting by $\gamma_\lambda^{(m)}$ the decay rates of the spins into modes propagating in directions $\lambda\!=\!L,R$ along waveguide $m$, the ME for multiple chiral waveguides reads
\begin{align}
\dot{\rho}=-\frac{i}{\hbar}\Big[H_{\rm sys}+\sum_{m,\lambda}\! H_{\lambda}^{(m)},\rho\Big]+\sum_{m,\lambda}\gamma_{\lambda}^{(m)}\mathcal{D}[c_\lambda^{(m)}]\rho.\label{MultiME}
\end{align}
Analogous to the single waveguide case, the coherent contributions from each waveguide $H_\lambda^{(m)}$, and the corresponding collective jump operators $c_\lambda^{(m)}$ are given by
\begin{align}
H_{\lambda}^{(m)}\!{}&\!\equiv\!\frac{-i\hbar\lambda\gamma_{\lambda}^{(m)}}{2}\!\sum_{j,l}\!\theta(x_j^m\!\!-\!x_l^m)\big(e^{ik|x_j^m-x_l^m|}\sigma_j\dg \sigma_l\!-\!{\rm h.c.}\big),\label{CoherentMulit}\\
c_\lambda^{(m)}&\equiv \sum_j e^{-ik\lambda x_j^m}\sigma_j.
\end{align}
Here we denoted the position of spin $j$ along waveguide $m$ by $x_j^m$ and assigned the values $\lambda=\{1,-1\}$ corresponding to $\lambda=\{R,L\}$. The symmetry breaking $\gamma_L^{(m)}\neq \gamma_R^{(m)}$ in the coupling to each reservoir, introduces an explicit dependence of the reservoir-mediated coherent term on the ordering of the spins along each waveguide, as reflected by the Heaviside-function $\theta(x)$ in Eq.~\eqref{CoherentMulit}, with $\theta(x)=1$ for $x>0$ and $\theta(x)=0$ for $x\leq0$.

These chiral networks allow coupling the spins in multiple ways, offering a variety of possibilities to create multipartite entangled states as we illustrate in the next subsection.

\subsection{Dynamical purification of spin multimers}\label{dynamical}

The master equations presented above describe the spin networks as a driven, open many-body system, whose dynamics drive the system into a steady state $\rho(t)\xrightarrow{t\rightarrow\infty}\rho_{ss}$. Generally, this steady state is mixed, but under special circumstances the interplay between driving and dissipation leads to a pure steady state $\rho_{ss}=\ket{\Psi}\bra{\Psi}$, which in the language of quantum optics is called a \emph{dark state} \cite{Parkins:1993ko}. There are a variety of paradigmatic examples of this in quantum optics, including optical pumping \cite{{HAPPER:1972fi}} and laser cooling \cite{metcalf}, where the internal or motional states of atoms, respectively, are dissipatively purified to a reach a steady state with a lower temperature. 

We will show below that for a spin ensemble coupled via a chiral network ($\gamma_L\neq\gamma_R$), there is a set of sufficient conditions under which the steady state is pure. In particular, for a single-waveguide network this set is as follows: 
\begin{enumerate}
\item[(i)] The spacing between spins is commensurate with the wavelength of reservoir excitations such that $k|x_j-x_l|=2\pi n$, with $n$ an integer.
\item[(ii)] All spins are driven symmetrically, $\Omega_j=\Omega$.
\item[(iii)] The total number of spins $N$ is \emph{even}.
\item[(iv)] The detuning pattern $\delta_j$ ($j=1,\dots, N$) is such that detunings cancel in pairs. That is, for each spin $j$ with detuning $\delta_j$, there is another spin $l$ with $\delta_l=-\delta_j$.  
\end{enumerate}
With conditions (i) and (ii) the chiral master equation \eqref{chiralME} can be written as
\begin{align}
\dot\rho=&-i\Big[\Omega(c+c^\dag)\!-\!\sum_j\delta_j\sigma_j^\dag \sigma_j\!-\!\frac{i\Delta\gamma}{2}\sum_{j>l}(\sigma_j^{\dag}\sigma_l-\sigma_l^\dag \sigma_j),\rho\Big]\nonumber\\
&+(\gamma_L+\gamma_R){\cal D}[c]\rho,\label{Dickebroken}
\end{align}
where condition (i) allows us to express the dissipative part in terms of a single collective jump operator $c\!=\!c_R\!=\!c_L\!=\!\sum_j\sigma_j$. In the bidirectional case, this condition leads to a complete absence of dipole-dipole interactions [cf.~Eq.~\eqref{purelydissipativeDicke}]. In \eref{Dickebroken}, however, owing to the asymmetry between decay rates $\Delta\gamma\equiv\gamma_R-\gamma_L>0$, spin-spin interactions are still present. Moreover, this chirality breaks the permutation symmetry between the spins, which is crucial for the \emph{uniqueness} of the steady state.

If conditions (iii) and (iv) are also fulfilled, the steady states of \eref{Dickebroken} are always pure and multipartite entangled, and their structure is determined by the detuning pattern $\delta_j$. In general, the steady state factorizes in a product of $N_m$ adjacent multimers:
\begin{align}\label{General_SS}
\ket{\Psi}=\bigotimes_{q=1}^{N_m}\ket{M_q},
\end{align}
where each multimer state $\ket{M_q}$ is a $M_q$-partite entangled state of an \emph{even} number of spins $M_q\leq N$. Specifically, it takes the form  
\begin{align}
\ket{M_q}&\!=\!a^{(0)}\ket{g}^{\otimes M_q}\!+\!\sum_{j_1<j_2}a^{(1)}_{j_1,j_2}\ket{S}_{j_1,j_2}\ket{g}^{\otimes M_q-2}\nonumber\\
&+\!\!\sum_{j_1<j_2<j_3<j_4}\!\!a^{(2)}_{j_1,j_2,j_3,j_4}\ket{S}_{j_1,j_2}\ket{S}_{j_3,j_4}\ket{g}^{\otimes M_q-4}\nonumber\\
&+\!\dots\!+\!\!\!\!\!\!\!\!\sum_{j_1<\dots<j_{M_q}}\!\!\!a^{(M_q/2)}_{j_1,\dots,j_{M_q}}\ket{S}_{j_1,j_2}\!\dots\!\ket{S}_{j_{M_q-1},j_{M_q}}\!,\label{multimer}
\end{align}
where $\ket{S}_{j,l}\equiv(\ket{g}_j\!\ket{e}_l-\ket{e}_j\!\ket{g}_l)/\sqrt{2}$ denotes the singlet state between two spins $j$ and $l$. These clusters contain superpositions of up to $M_q/2$ (delocalized) singlets, but no other spin excitations.
\begin{figure}[b]
\includegraphics[width=0.5\textwidth]{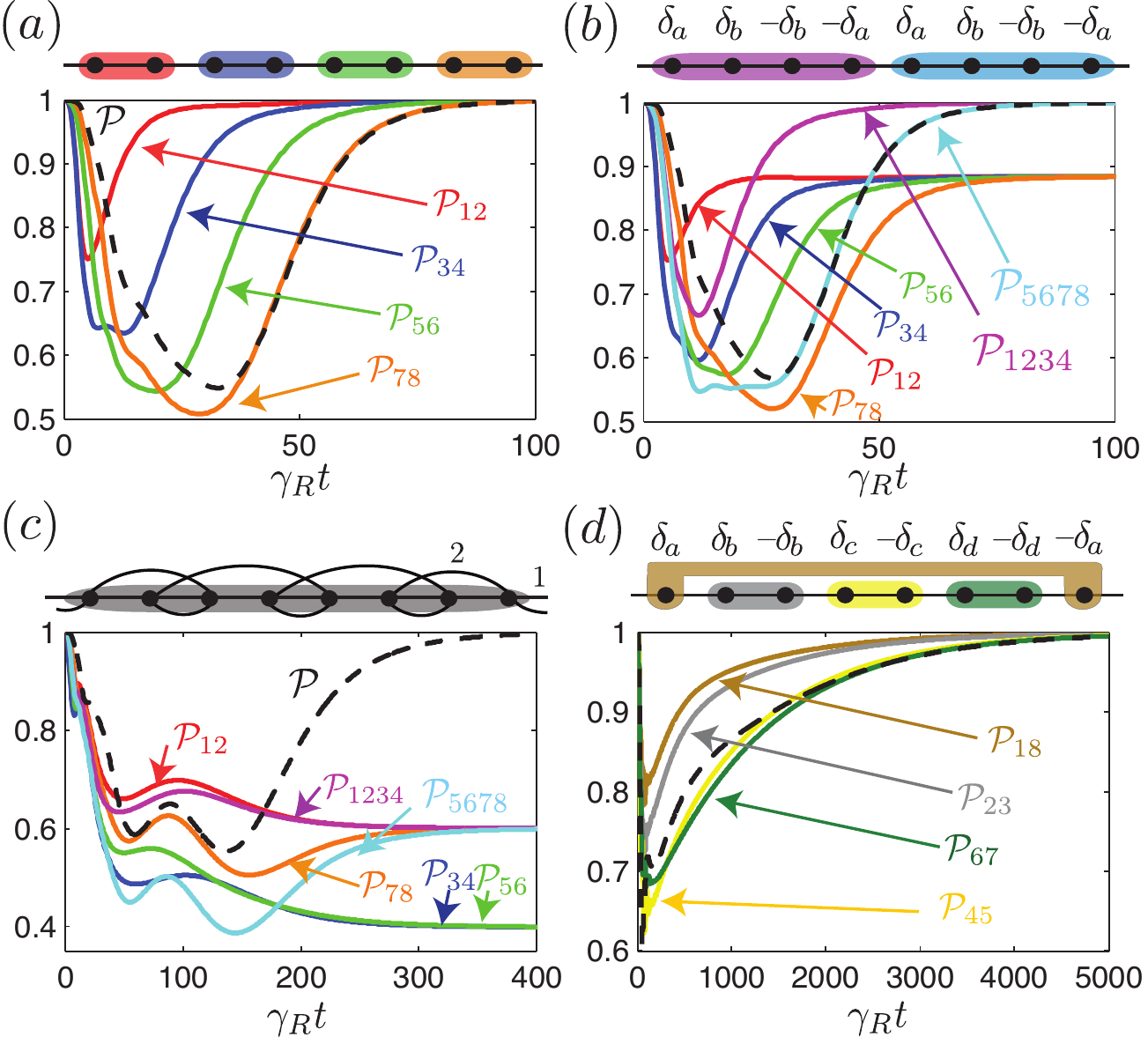}
\caption{(Color online) Dynamical purification of a chiral spin network of $N\!=\!8$ spins into different entangled multimer steady states. As a function of time, we plot the purity of the total state $\cal P$ and the purities of reduced density matrices of different spin subsets ${\cal P}_{j_1,j_2,\dots}\equiv\!{\rm Tr}\{(\rho_{j_1,j_2,\dots})^2\}$ to probe the entanglement structure of the steady states. (a) Dimers are formed when $\delta_j=0$ and $\gamma_L=0.1\gamma_R$. (b) Tetramers are formed for the indicated detuning pattern. Here $\delta_a\!=\!0$, $\delta_b\!=\!0.3\gamma_R$ and $\gamma_L\!=\!0.1\gamma_R$. (c) Genuine 8-partite entangled octamer formed as the result of coupling the spins to two chiral channels, when driven on resonance $\delta_j=0$. For the second chiral channel we assume $\gamma_R^{(2)}\!=\!\gamma_R$, $\gamma_L^{(2)}\!=\!0.5\gamma_R$ and the order of coupling the spins is indicated above. Additionally, $\gamma_L^{(1)}\!=\!0.1\gamma_R$ and $\gamma_R^{(1)}\!=\!\gamma_R$. (d) Non-local dimers in a single bidirectional channel, $\gamma_L\!=\!\gamma_R$. The detuning pattern is as indicated, with $\delta_a\!=\!0.6\gamma_R$, $\delta_b\!=\!0.4\gamma_R$, $\delta_c\!=\!0.2\gamma_R$, and $\delta_d\!=\!0.1\gamma_R$. All calculations assume $\Omega\!=\!0.5\gamma_R$.}\label{fig:purification_teaser}
\end{figure}

As an illustrative example, in Fig.~\ref{fig:purification_teaser} we analyze the dynamics that produce entangled pure states in a chiral network of $N\!=\!8$ spins. This will be expanded upon in Secs.~\ref{steadystates} and \ref{Nspins}, where we analyze in detail the general conditions leading to these types of states. In Fig.\,\ref{fig:purification_teaser}(a) all spins are driven on resonance $\delta_j\!=\!0$, which results in the spin chain dynamically purifying into \emph{dimers} (i.e.,~$M_q=2$, $\forall q$). Here, not only the total state purifies dynamically [${\cal P}(t)\!\equiv\!{\rm Tr}\{\rho^2(t)\}\rightarrow 1$ as $t\rightarrow\infty$], but also the reduced states of spin pairs [${\cal P}_{2j-1,2j}\!\equiv\!{\rm Tr}\{(\rho_{2j-1,2j})^2\}\rightarrow 1$, as $t\rightarrow\infty$, $\forall j=1,\dots,4$]. Throughout this work we use the notation $\rho_{j_1,j_2,\dots}$ to denote the reduced density operator of spins $\{j_1,j_2,\dots\}$. On the other hand, when driving the same chiral spin network with the detuning pattern indicated in Fig.\,\ref{fig:purification_teaser}(b), the spin chain arranges itself into four-partite clusters, or \emph{tetramers}, as indicated by the purification of two blocks of four adjacent spins [cf.~ Fig.~\ref{fig:purification_teaser}(b)]. All other spin subsets are in a mixed steady state. 

It is remarkable that also in a multiple-waveguide setting, pure states of the form in Eq.~\eqref{General_SS} can be obtained under analogous conditions as (i)-(iv) (see Sec.\,\ref{multiwaveguideN} for details). In Fig.~\ref{fig:purification_teaser}(c) we show an example with $N\!=\!8$ spins coupled to two chiral waveguides in a partially reverse order, as indicated. When driven on resonance $\delta_j\!=\!0$, the spins purify into a genuine $8$-partite entangled state, or $\emph{octamer}$. In this case only the total state purifies ${\cal P}(t)\rightarrow 1$, while all reduced density matrices involving fewer spins stay mixed. While a chiral spin chain with a single waveguide forms a dimerized state when driven on resonance, the alternate ``wiring'' of the second waveguide is here the key to entangle these structures.

We note that also in the case of an ideal bidirectional waveguide ($\gamma_L\!=\!\gamma_R$) it is possible to prepare unique pure steady states in specific situations. While in the chiral setting the permutation symmetry is intrinsically broken via the chirality of the reservoir, in the bidirectional case this can also be achieved by choosing different detunings for each spin. Then, under the same conditions as in the chiral case (i)-(iv), the system is driven into a unique steady state. However, only bipartite dimerized states form, but interestingly, depending on the detuning pattern, these can be highly non-local. For instance, in Fig.~\ref{fig:purification_teaser}(d) we illustrate such a situation. There, the first and last spins are driven into a non-local pure entangled state, while all the other spins in between are dynamically purified into adjacent dimers. Note that in the absence of chirality, the coupling between subspaces of different permutation symmetry is weaker, and correspondingly the time scale to approach this steady state is longer than in the chiral counterpart [cf.~Fig.~\ref{fig:purification_teaser}(b)].

\subsection{Experimental realizations}

\begin{figure}[t!]
\includegraphics[width=0.44\textwidth]{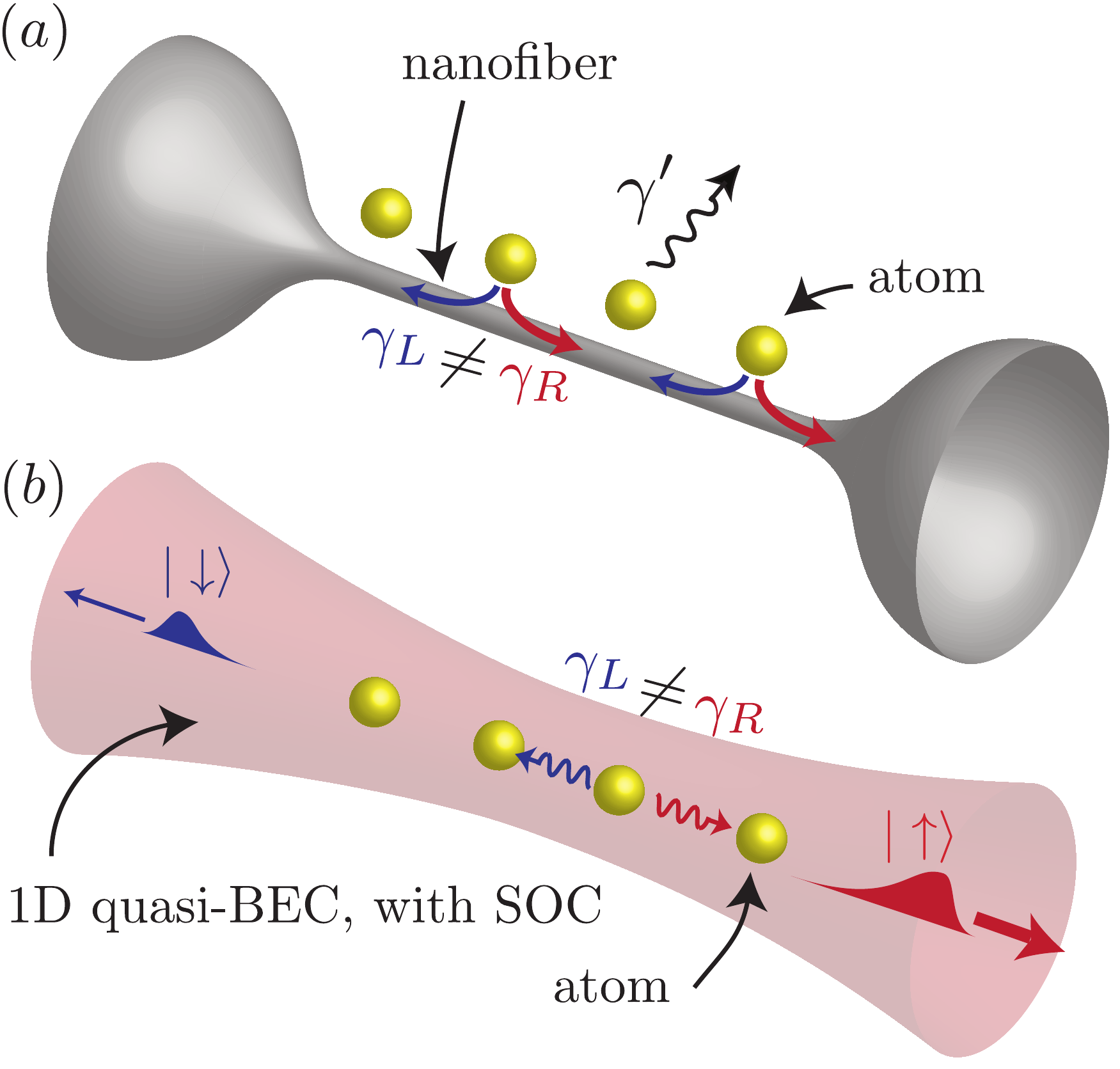}
\caption{(Color online) Photonic and phononic realizations of spin chains with chiral coupling to a 1D reservoir. (a) Atoms coupled to the guided modes of a tapered nanofiber. The directionality of the photon emission  $\gamma_L\neq\gamma_R$ stems from coupling between the transverse spin density of light and its propagation direction. A current experimental challenge is the control of photon emission into non-guided modes, indicated by $\gamma'$. (b) Cold atoms in an optical lattice immersed in a 1D quasi-BEC of a second species of atoms, where the latter represents the reservoir \cite{{Chen:2014it},{Scelle:2013in}}. Including synthetic spin-orbit coupling (SOC) of the quasi-BEC \cite{Lin:2011hn} allows the breaking of the symmetry of decay into left and right moving Bogoliubov excitations \cite{Ramos:2014ut}.}
\label{fig:chirality}
\end{figure}

As a final remark in this overview section, it is important to comment on experimental possibilities to realize these chiral spin networks. Very recently, chiral system-reservoir interactions of the type in Eq.~(\ref{chiralInt}) have been realized in photonic systems by coupling a Cs atom \cite{{Mitsch:2014tr}} or a gold nanoparticle \cite{Petersen:2014tc} to a tapered nanofiber as shown in Fig.~\ref{fig:chirality}(a), as well as quantum dots to photonic crystal waveguides \cite{Sollner:2014wb}. Related works where directionality of photon emission has been experimentally demonstrated or proposed can be found in Refs.~\cite{Neugebauer:2014iy,RodriguezFortuno:2013ew,Lin:2013do,Young:2014ty}. 
The directionality in these photonic setups is due the strong transverse confinement of light, which gives rise to non-paraxial longitudinal components of the electric field that are different for left and right moving photons \cite{Bliokh:2014cd,Neu2015aa}. A polarization-selective coupling of an emitter can therefore result in directional coupling to the guided modes \cite{{Mitsch:2014tr},Petersen:2014tc,Sollner:2014wb}. Note that also optomechanical systems have been proposed to realize a unidirectional spin chain \cite{Stannigel:2012jk}.

On the other hand, a purely atomic implementation of these chiral reservoirs, replacing photons by phonons, has been proposed in Ref.~\cite{Ramos:2014ut} [cf.~Fig.~\ref{fig:chirality}(b)]. There, cold atoms in an optical lattice are immersed in a second species of atoms representing the reservoir \cite{{Chen:2014it},{Scelle:2013in}}. This reservoir gas is confined to a 1D geometry and forms a quasi-BEC in which the Bogoliubov excitations play the role of the guided modes. The symmetry between left and right moving modes is broken by implementing synthetic spin-orbit coupling of the reservoir gas \cite{{Lin:2011hn},Galitski:2013dh}, as detailed in Ref.~\cite{Ramos:2014ut}. 
Proof of principle experiments on implementations of such quantum optical systems with cold atoms have already been reported in Refs.~\cite{{Chen:2014it},{Scelle:2013in}}.
One of the remarkable features of this implementation is the intrinsic absence of other decay channels outside the waveguide [cf.~Fig.~\ref{fig:chirality}(a)], which is currently a major challenge in photonic experiments.

\section{Pure dark steady states of chiral spin networks}\label{steadystates}

From a quantum optics perspective, steady states of open systems are pure when they are \emph{dark states} of the driven-dissipative dynamics. The scope of this section is to analyze in detail the conditions under which the steady states of the chiral spin networks are dark, and to establish a physical interpretation of the underlying mechanisms. In particular, in the illustrative examples with two and four spins, we show that the conditions stated in Sec.\,\ref{dynamical} are sufficient to cool the system into such dark states. In Sec.\,\ref{Nspins} we extend this to larger networks.

We recall that a pure quantum state $\ket{\Psi}$ is a dark state of the driven-dissipative dynamics \cite{{Kraus:2008jd},{Baumgartner:2008ic}}, if it is
\begin{enumerate}
\item[(1)] annihilated by all jump operators, and 
\item[(2)] invariant under the coherent part of the dynamics, i.e.~an eigenstate of the Hamiltonian. 
\end{enumerate} 
In the particular case of a chiral spin network with ME \eqref{chiralME}, the first condition reads $c_L\ket{\Psi}\!=\!c_R\ket{\Psi}\!=\!0$, which means that the system does not emit photons at both output ports of the waveguide (hence the term ``dark''). The second condition is fulfilled if $(H_{\rm sys}+H_{L}+H_{R})\ket{\Psi}=E\ket{\Psi}$,~i.e., if the state is an eigenstate of the total Hamiltonian, consisting not only of the system part $H_{\rm sys}$, but also of the bath-mediated coherent parts $H_{L}$ and $H_{R}$. In general, these two conditions can not be satisfied at the same time, inhibiting the existence of a dark state. To understand why and when they can be satisfied simultaneously it is instructive to first consider the simple example of only two spins coupled by a chiral waveguide, since it contains many of the essential features, and will serve as a building block to understand larger systems.

\subsection{Two spins coupled by a chiral waveguide}\label{two_spins}

The dark state condition (1) restricts the search for dark states to the null spaces of the two jump operators $c_L$ and $c_R$. The null space of $c_L\!=\!\sigma_1\!+\!e^{ik|x_2-x_1|}\sigma_2$ is spanned by the trivial state $\ket{gg}$ and the state $\ket{\Psi_L}\!\equiv\!(\ket{ge}-e^{ik|x_2-x_1|}\ket{eg})/\sqrt{2}$. The latter does not emit photons propagating to the left because of destructive interference of the left-moving  photons emitted by the two spins, an effect well known as subradiance \cite{Gross:1982js,Lehmberg:1970kn}. However, this subradiant state in general decays by emitting photons traveling to the right;~i.e., $c_R\ket{\Psi_L}\neq 0$. On the other hand, the null space of $c_R\!=\!\sigma_1+e^{-ik|x_2-x_1|}\sigma_2$ is spanned by $\ket{gg}$ and $\ket{\Psi_R}\!\equiv\!(\ket{ge}\!-\!e^{-ik|x_2-x_1|}\ket{eg})/\sqrt{2}$, where again, $\ket{\Psi_R}$ is subradiant with respect to emission of photons to the right, but in general emits photons to the left. As a consequence, only the trivial state $\ket{gg}$ is in general annihilated by both jump operators, leaving no room for a nontrivial dark state.
\begin{figure}[t]\centering
\includegraphics[width=0.50\textwidth]{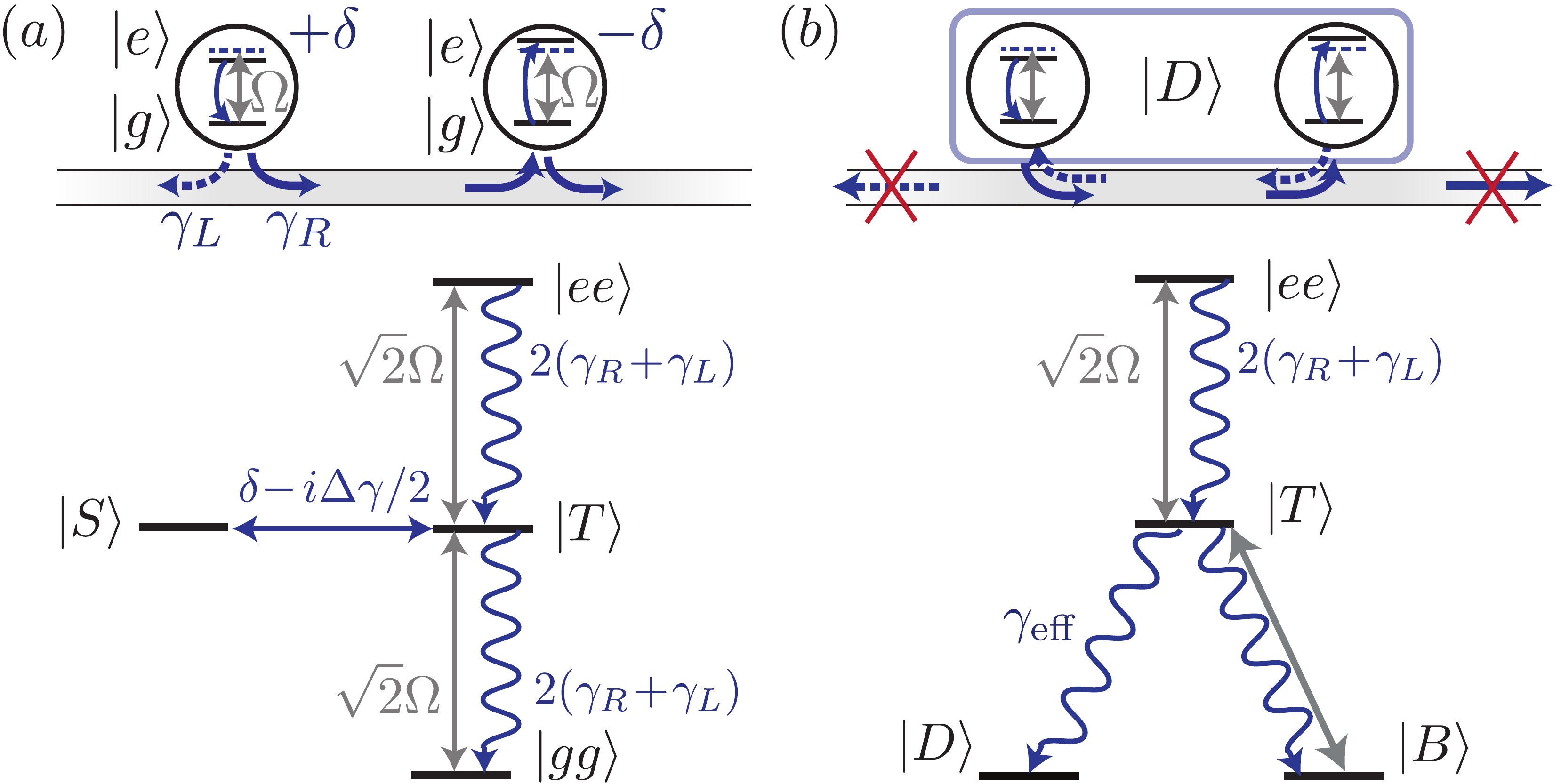}
\caption{(Color online) A chiral waveguide couples two spins that are separated by a distance commensurate with the photon wavelength. They are additionally driven homogeneously ($\Omega_1\!=\!\Omega_2\!=\!\Omega$) and with opposite detunings ($\delta_1\!=\!-\delta_2\!=\!\delta$). (a) The superradiant collective decay couples dissipatively only the spin triplet states. The spin singlet $\ket{S}$ does not emit into the waveguide (subradiance) and couples coherently only to $\ket{T}$. (b) The level diagram of states $\ket{gg}$, $\ket{S}$, and $\ket{T}$ resembles a $\Lambda$ system and thus there is a dark state $\ket{D}$ in the subspace spanned by $\ket{S}$ and $\ket{gg}$.}
\label{fig:2spins_ideal}
\end{figure}
An exception occurs if the distance of the two spins is an integer multiple of the wavelength of the photons, that is if $k|x_1-x_2|\!=\!2\pi n$ with $n=0,1,2,\dots$ \cite{footnote1}. Then the two jump operators coincide $c_L\!=\!c_R\!=\!c\!=\sigma_1+\sigma_2$ (up to an irrelevant phase), and the common null space is spanned by the two states $\ket{gg}$ and $\ket{S}\equiv(\ket{ge}-\ket{eg})/\sqrt{2}$. The so-called singlet state $\ket{S}$ is then perfectly subradiant with respect to both photons propagating to the right and photons propagating to the left. On the other hand the triplet state $\ket{T}\equiv(\ket{ge}+\ket{eg})/\sqrt{2}$ is superradiant; that is, it decays with $2(\gamma_L+\gamma_R)$ (cf.~Fig.\,\ref{fig:2spins_ideal}). We note that in the perfectly cascaded setup this condition on the distance of the spins is not required, since in this case there is only one jump operator [cf.~Eq.~(\ref{Cascaded_ME})]. 

\begin{figure}[b]\centering
\includegraphics[width=0.48\textwidth]{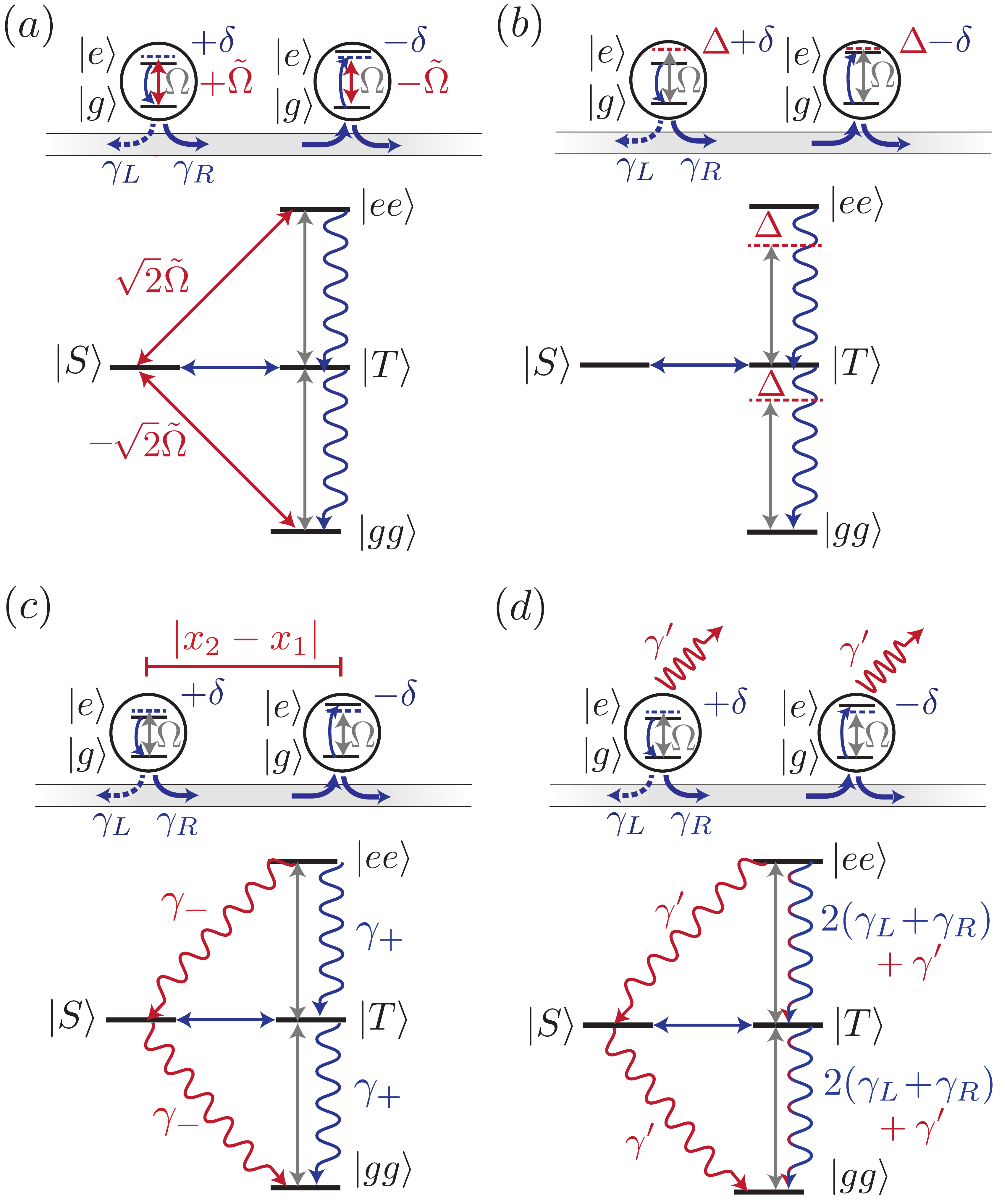}
\caption{(Color online) Deviations from the dark state conditions for $N=2$ and their effect on the dark states. (a) A nonhomogeneous drive $\tilde\Omega\!=\!\Omega_1\!-\!\Omega_2\!\neq\!0$ couples $\ket{S}$ coherently to $\ket{ee}$ and $\ket{gg}$, inhibiting the formation of a dark state. (b) A homogeneous offset in the detunings $\Delta\!=\!(\delta_1\!+\!\delta_2)/2\!\neq\!0$ destroys the Raman resonance between the states $\ket{S}$ and $\ket{gg}$. (c) Decay processes for spins at arbitrary distance,~i.e., noncommensurate with the wavelength. The state $\ket{S}$ is in general not perfectly subradiant and thus decays to the state $\ket{gg}$. (d) Additional decay channels such as emission of the spins into independent reservoirs different from the chiral waveguide also lead to a decay of the singlet state.}
\label{fig:2spins_imperfection}
\end{figure}
As mentioned in Sec.~\ref{dynamical}, condition (2) can be satisfied if the two spins are driven with the same Rabi frequency $\Omega_1=\Omega_2\equiv\Omega$ and opposite detunings $\delta_1=-\delta_2\equiv\delta$. This can be most easily realized by expressing the Hamiltonian $H=H_{\rm sys}+H_{L}+H_{R}$ in the basis of singlet $\ket{S}$ and the triplet states $\ket{gg}$, $\ket{T}$, and $\ket{ee}$ as 
\begin{align}
\frac{H}{\hbar}\!=\!\sqrt{2}\Omega\!\lr{\ket{T}\!\!\bra{gg}\!+\!\ket{ee}\!\!\bra{T}}\!+\!\!\lr{\!\delta\!-\!\frac{i\Delta\gamma}{2}\!}\!\!\ket{S}\!\!\bra{T}\!+\!\rm h.c.
\end{align}
The level scheme and the corresponding coherent and dissipative couplings of the two spins are depicted in Fig.~\ref{fig:2spins_ideal}(a). There the states $\ket{gg}$, $\ket{S}$, and $\ket{T}$ resemble a so-called $\Lambda$ system with resonant couplings from the stable states $\ket{gg}$ and $\ket{S}$ to the superradiant state $\ket{T}$. In the null space of $c$, one can thus find a transformation from $\ket{gg}$ and $\ket{S}$ to a dark state $\ket{D}$ and a bright state $\ket{B}$ [cf.~Fig.~\ref{fig:2spins_ideal}(b)]. The state $\ket{D}$ decouples from the coherent dynamics due to destructive interference of the coherent drive, chiral interactions, and detunings, and thus is an eigenstate of $H$. Explicitly it is given by
\begin{align}\label{Dimer}
\ket{D(\alpha)}\equiv\frac{1}{\sqrt{1+|\alpha|^2}}\lr{\ket{gg}+\alpha\ket{S}},
\end{align}
where $\alpha\equiv -2\sqrt{2}\Omega/(2\delta+i\Delta\gamma)$ is the \emph{singlet fraction}. Figure~\ref{fig:2spins_ideal}(b) shows the coherent and dissipative couplings in this transformed basis. The two spins are dissipatively ``pumped'' into this dark state on a timescale $t_D\equiv 2\pi/\gamma_{\rm eff}$, where the effective decay $\gamma_{\rm eff}$ reads
\begin{align}\label{Timescale_dimer}
\gamma_{\rm eff}=\frac{2(\gamma_L+\gamma_R)(\Delta\gamma^2/4+\delta^2)}{\Delta\gamma^2/4+\delta^2+2\Omega^2}.
\end{align}

\begin{figure}[t]
\includegraphics[width=0.50\textwidth]{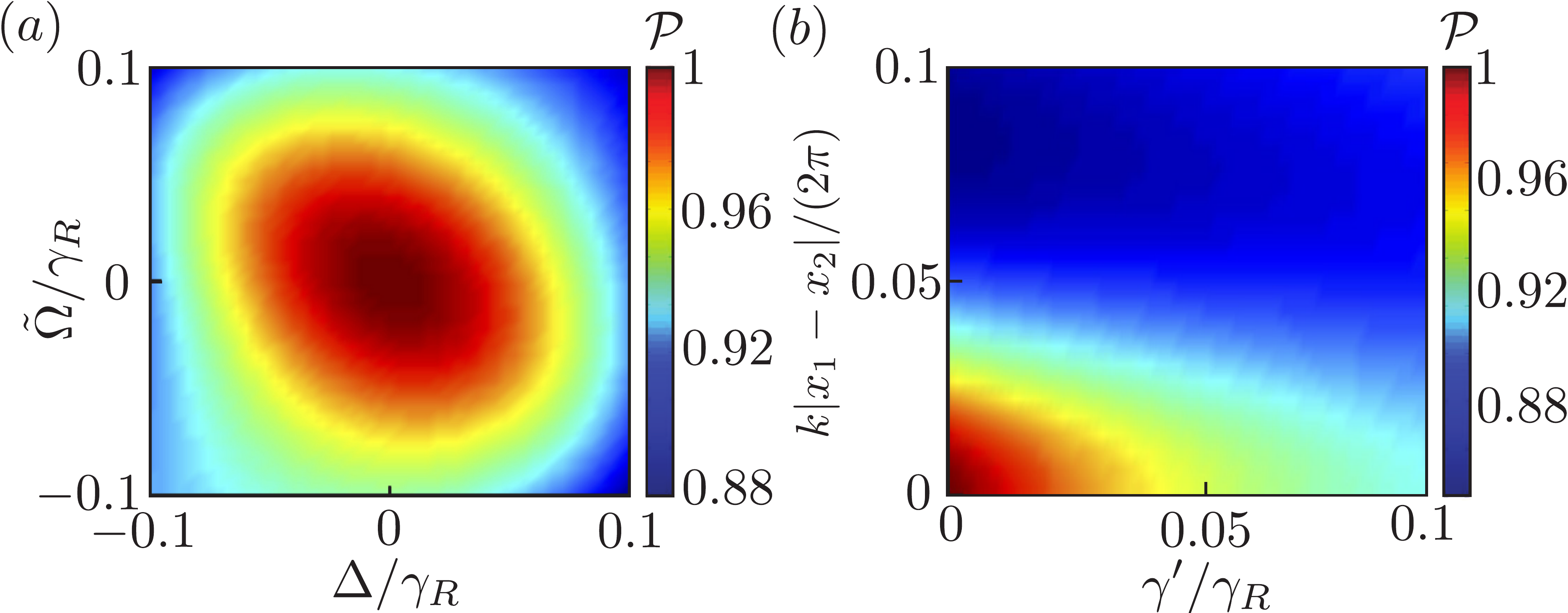}
\caption{(Color online) Purity of the steady state for $N=2$ if the dark-state conditions are not met (cf.~Fig.\,\ref{fig:2spins_imperfection}). (a) Shown as a function of a homogeneous offset in the detuning $\Delta$ and a staggered component of the coherent drive $\tilde\Omega$. (b) As a function of the distance between the spins relative to the wavelength (modulo integers) and an on-site decay $\gamma'$. Parameters are $\Omega/\gamma_R=0.5$, $\delta/\gamma_R=0.3$, and $\gamma_L/\gamma_R=0.5$.}\label{fig:imperfection_2spins}
\end{figure}
With this picture it is simple to understand the two requirements $\Omega_1\!=\!\Omega_2$ and $\delta_1\!=\!-\delta_2$, necessary for the existence of a pure steady state. An inhomogeneous Rabi frequency leads to a coherent coupling of the singlet state $\ket{S}$ to the states $\ket{gg}$ and $\ket{ee}$ with strength $\sqrt{2}\tilde\Omega$, where $\tilde\Omega\!\equiv\!\Omega_1\!-\!\Omega_2$ [cf.~Fig.\,\ref{fig:2spins_imperfection}(a)], inhibiting the existence of an eigenstate of the Hamiltonian in the null space of $c$. On the other hand, if the detunings are not exactly opposite, but rather have an additional homogeneous offset $\Delta\!\equiv (\delta_1\!+\!\delta_2)/2\neq 0$, this gives rise to a nonzero ``two-photon'' Raman detuning between states $\ket{S}$ and $\ket{gg}$. Effectively this leads to a coupling between the ideal dark state $\ket{D}$ and the bright state $\ket{B}$, inhibiting the formation of a stationary dark state [cf.~Fig.~\ref{fig:2spins_imperfection}(b)]. Similarly, also other imperfections are simple to understand in this picture. For example we show in Fig.\,\ref{fig:2spins_imperfection}(c) the dissipative couplings if the commensurability condition (i) is not met. As discussed above, the singlet is no longer in the null space of both jump operators $c_L$, $c_R$ and decays with a rate $\gamma_{-}$, where we denoted $\gamma_{\pm}\equiv(\gamma_L+\gamma_R)[1\pm\cos(k|x_2-x_1|)]$. In addition, an experimentally relevant question is the effect of a finite decay to dissipative channels other than the chiral waveguide \cite{Mitsch:2014tr,Petersen:2014tc}. This would introduce an additional term 
$\gamma'(\mathcal{D}[\sigma_1]\rho+\mathcal{D}[\sigma_2]\rho)$ to the chiral ME \eqref{chiralME}. Since the singlet $\ket{S}$ is not in the null spaces of these two additional jump operators $\sigma_1$ and $\sigma_2$, the pure dark state does not survive [cf.~Fig.\,\ref{fig:2spins_imperfection}(d)]. In Fig.~\ref{fig:imperfection_2spins} we quantify the decrease in the purity $\mathcal{P}\!\equiv\!\textrm{Tr}\{\rho_{ss}^2\}$ of the steady state of the chiral ME (\ref{chiralME}), when considering different deviations of the dark-state conditions discussed above. In general, one finds that the steady state is close to pure if these deviations are small compared to the rate $\gamma_{\rm eff}$ at which the dark steady state would be formed in the ideal setup, i.e., $\tilde{\Omega},\Delta,\gamma_{-},\gamma'\ll\gamma_{\rm eff}$.

As a final remark in this subsection we note that the dimers are formed as long as the permutation symmetry between the two spins is broken,~i.e., as long as $\Delta\gamma\neq 0$ or $\delta\neq 0$. Else the singlet and the triplet manifold decouple [cf.~Fig.~\ref{fig:2spins_ideal}(a)] and the steady state is not unique.

\subsection{Four spins coupled to a chiral waveguide}\label{N4}

To gain insight into the general structure of dark states of longer spin chains, we consider here the case of $N=4$ spins. This system is still small enough to find analytical solutions and allows us to show that conditions (i)-(iv) are necessary and sufficient to obtain dark states (up to trivial exceptions). Moreover, it will pave the way to the more general discussion of larger networks in Sec.~\ref{Nspins}. Note that $N\!=\!3$ spins do not allow for pure dark states, as direct search shows.

\begin{figure}[t]
\includegraphics[width=0.50\textwidth]{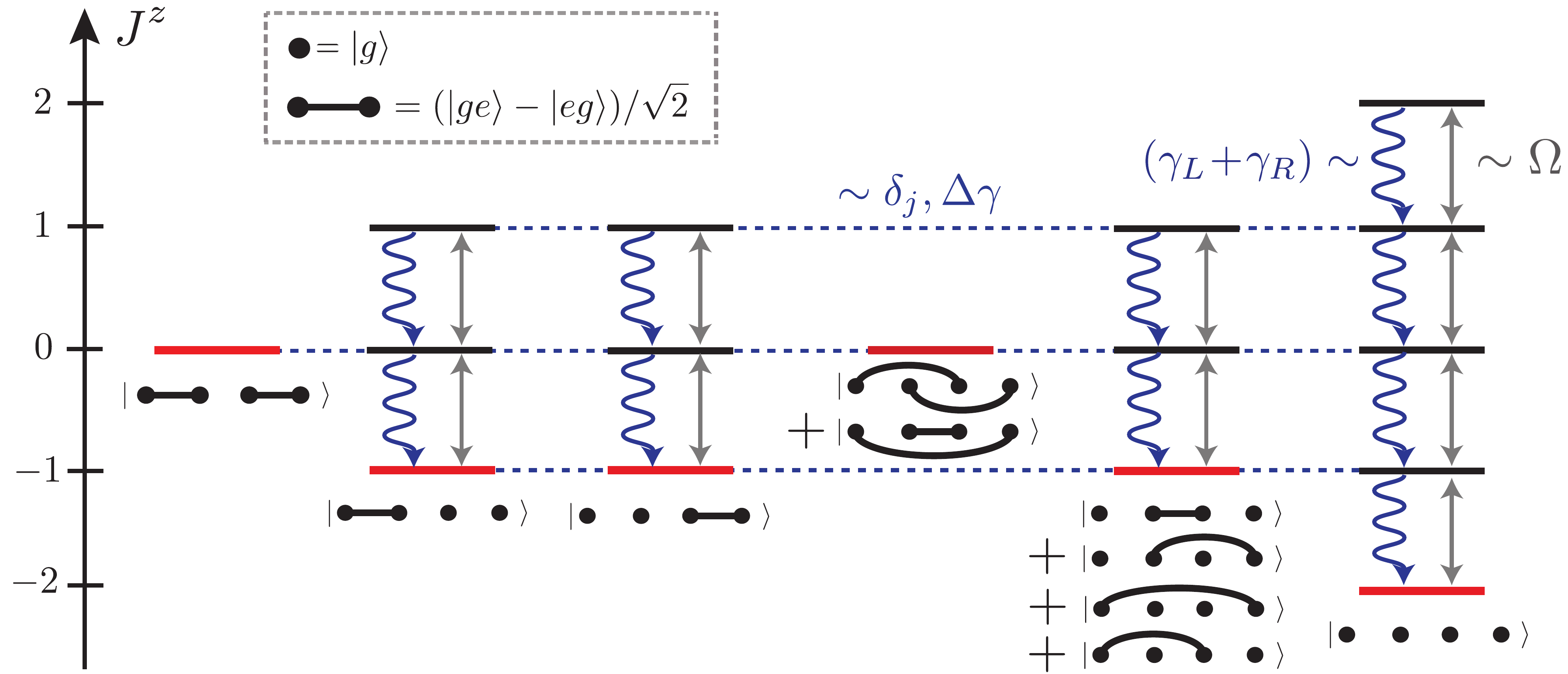}
\caption{(Color online) Level diagram of the $N\!=\!4$ spin system in a total angular momentum basis, obtained by first adding the subspaces of spin 1 with 2 and spin 3 with 4, separately. The resulting $16$ states are grouped into 6 angular momentum manifolds of given total angular momentum, which ranges from 0 to 2 (see text). In each manifold, states are ordered by increasing number of excited spins, i.e., eigenvalue of $J^z$ (see text). The coherent driving $\sim\!\Omega$ and dissipative terms $\sim(\gamma_L\!+\!\gamma_R)$ couple them vertically. The interactions $\Delta\gamma$ and different detunings $\sim\!\delta_j$ couple states of different manifolds, but conserve the number of excitations. The null space of the collective jump operator $c=\sum_j \sigma_j$ is spanned by the 6 states marked in red. All these are superpositions of products of singlet $\ket{S}_{j,l}\!\equiv\!(\ket{ge}\!-\!\ket{eg})/\sqrt{2}$ and $\ket{g}_j\ket{g}_l$ states between the different spins $j,l$, as indicated in the figure.}\label{fig:tetramer_levels}
\end{figure}

As in Sec.~\ref{two_spins} we start by identifying the null space of the jump operators $c_L$ and $c_R$. Again, its dimension depends on the distances between the spins with respect to $k$, and is maximal if both jump operators coincide $c_L=c_R=c$, that is, if the commensurability condition (i) of Sec.\,\ref{dynamical} is fulfilled.
The corresponding null space is then spanned by states in which excitations of the spins are always shared in singlet states $\ket{S}_{j,l}$ between two spins $j$ and $l$, while all other spins are in the state $\ket{g}$. Therefore, condition (1) restricts the possible dark states to states of the form [cf.~Eq.\,(\ref{multimer})]
\begin{align}
\ket{\Psi}=&\ a^{(0)}|gggg\rangle\!+\!a^{(1)}_{12}|S\rangle_{12}|gg\rangle_{34}\!+\!a^{(1)}_{34}|S\rangle_{34}|gg\rangle_{12}\nonumber\\
&\!+\!a^{(1)}_{13}|S\rangle_{13}|gg\rangle_{24}\!+\!a^{(1)}_{14}|S\rangle_{14}|gg\rangle_{23}\!+\!a^{(1)}_{23}|S\rangle_{23}|gg\rangle_{14}\nonumber\\
&\!+\!a^{(1)}_{24}|S\rangle_{24}|gg\rangle_{13}\!+\!a^{(2)}_{1234}|S\rangle_{12}|S\rangle_{34}\nonumber\\
&\!+\!a^{(2)}_{1324}|S\rangle_{13}|S\rangle_{24}\!+\!a^{(2)}_{1423}|S\rangle_{14}|S\rangle_{23}.\label{explicit4state}
\end{align}
One can easily check that this subspace for four spins is six-dimensional. Note that any violation of the commensurability condition leads to a second (independent) jump operator, which reduces the dimension of this subspace, inhibiting in general a simultaneous fulfillment of the dark-state conditions (1) and (2). To see that states of the form in Eq.~\eqref{explicit4state} indeed span the full null space of $c$ it is instructive to add the four different spins-1/2 to a total angular momentum. For example, one can add the first two spins to form a spin-0 and a spin-1 system as in the $N=2$ case, and analogously the last two spins. Adding these spins, one obtains two spin-0, three spin-1, and one spin-2 system (cf.~Fig.~\ref{fig:tetramer_levels}). Note that the collective jump operator $c=\sum_j\sigma_j$ is simply the lowering operator of the collective angular momentum, such that in each of these six manifolds there is exactly one state, namely the one with the minimum eigenvalue of the $z$ component of the total angular momentum $J^z\!\equiv\!\sum_j(\sigma_j^\dag\sigma_j\!-\!\sigma_j\sigma_j^\dag)$, which is annihilated by $c$ (cf.~red states in Fig.\,\ref{fig:tetramer_levels}). 

Dark states can be formed if $H_{\rm sys}+H_L+H_R$ has an eigenstate in this null space. As in the $N\!=\!2$ case, this happens only when all spins are driven homogeneously $\Omega_j\!=\!\Omega$, implying that the driving terms $\sim\!\Omega$ couple only states within the same angular momentum manifold (cf.~vertical arrows in Fig.\,\ref{fig:tetramer_levels}). On the other hand, the reservoir-mediated spin-spin interactions $\sim\Delta\gamma$, as well as differences in detunings $\sim\delta_j$, couple only states with the same number of excitations (cf.~horizontal dashed lines in Fig.\,\ref{fig:tetramer_levels}). 
It is a straightforward calculation to show that the existence of dark states, that is, eigenstates of $H_{\rm sys}+H_L+H_R$ in the six-dimensional null space of $c$, requires the detunings $\delta_j$ to vanish in pairs. For $N=4$ spins there are three different possibilities to satisfy this:
\begin{enumerate}
\item[(I)] $\delta_1+\delta_2=\delta_3+\delta_4=0$,
\item[(II)] $\delta_1+\delta_3=\delta_2+\delta_4=0$,
\item[(III)] $\delta_1+\delta_4=\delta_2+\delta_3=0$.
\end{enumerate}
The structure of the steady state thereby depends on the structure of this detuning pattern. Note that all three cases can be obtained from each other by permutations of the detunings.  
\subsubsection{Dimerization}
If the detuning pattern is of the form (I), one finds that the dark state decouples into two dimers:
\begin{align}
\ket{\Psi}=\ket{D(\alpha_1)}_{12}\ket{D(\alpha_3)}_{34}.\label{twodimerss}
\end{align}
The dimers $\ket{D(\alpha_1)}_{12}$ and $\ket{D(\alpha_3)}_{34}$, formed between between the first and the second spinpair, respectively, are of the same form as Eq.\,(\ref{Dimer}), with singlet fractions $\alpha_j$ defined as
\begin{align}
\alpha_j\equiv\frac{-2\sqrt{2}\Omega}{2\delta_j+i\Delta\gamma}.\label{singletfraction}
\end{align}
As is evident from Eq.\,(\ref{twodimerss}), under these conditions the dark state is two-partite entangled. This dimerized state is the straightforward generalization of the $N\!=\!2$ case presented in Sec.~\ref{two_spins}. Each of the two spin pairs thereby goes separately into a dark state, scattering no photons into the waveguide, and thus allowing also the other pair to reach its corresponding dark state. In the next section we show that this concept generalizes to chains with any even number of spins $N$.

\subsubsection{Tetramer}\label{tetramersubsec}

If the detuning pattern is not of the form (I) but fulfils (II) or (III), the dark state is fully four-partite entangled. The corresponding \emph{tetramer} states, which are special cases of the general multimer in Eq.\,(\ref{multimer}), read
\begin{align}
\ket{\Psi}\!&\propto\!|gggg\rangle\!+\!a^{(1)}_{12}|S\rangle_{12}|gg\rangle_{34}\!+\!a^{(1)}_{34}|S\rangle_{34}|gg\rangle_{12}\nonumber\\
&\!\!\!\!\!\!\!+\!a^{(1)}_{13}\!\big(|S\rangle_{13}|gg\rangle_{24}\!+\!|S\rangle_{14}|gg\rangle_{23}\!+\!|S\rangle_{23}|gg\rangle_{14}\!+\!|S\rangle_{24}|gg\rangle_{13}\big)\nonumber\\
&\!\!\!\!\!\!\!+\!a^{(2)}_{1234}|S\rangle_{12}|S\rangle_{34}\!+\!a^{(2)}_{1324}(|S\rangle_{13}|S\rangle_{24}\!+\!|S\rangle_{14}|S\rangle_{23}).\label{tetramer}
\end{align}
where the explicit form of the coefficients $a^{(m)}_{j_1,\dots,j_{2m}}\!\propto\!\Omega^m$ is given in Appendix \ref{tetramersolution}. Interestingly, in the strong-driving limit $\Omega\rightarrow\infty$, this tetramer takes the form $\ket{\Psi}\!\propto\!|S\rangle_{12}|S\rangle_{34}+|S\rangle_{13}|S\rangle_{24}+|S\rangle_{14}|S\rangle_{23}$, reminiscent of a valence bond state \cite{Auerbach:1994jf}.

\subsubsection{Nonlocal dimers in bidirectional bath}\label{nonlocaldimers}

The formation of such a tetramer relies heavily on the broken symmetry between $\gamma_L$ and $\gamma_R$. In fact, if $\Delta\gamma=0$, no four-partite entangled state can be formed in the dissipative dynamics. As already seen in the $N=2$ case, unique pure steady states exist also in the bidirectional case (under the same conditions as in the chiral case), if the permutation symmetry is broken via $\delta_j\neq 0$. However, a direct calculation (cf.~Appendix~\ref{tetramersolution}) shows that they are always dimerized. Remarkably, when the detuning pattern is of the form (II) or (III) [but not of the form (I)], these dimers are \emph{nonlocal} since the dark state factorizes as $\ket{\Psi}\!=\!\ket{D(\alpha_1)}_{13}\ket{D(\alpha_2)}_{24}$ or $\ket{\Psi}\!=\!\ket{D(\alpha_1)}_{14}\ket{D(\alpha_2)}_{23}$, respectively. In these two last cases pairs of non-neighboring spins are entangled, but they decouple from adjacent spins due to quantum interference. In Fig.\,\ref{fig:purification_teaser}(d) we show an example of this behavior in the case of $N=8$, where spins 1 and 8 form a nonlocal dimer in steady state. In Sec.~\ref{Nnonlocalpairs} we discuss this in the general context of networks with arbitrary even $N$. 

\section{Pure steady states in systems with $N$ spins}\label{Nspins}

In this section we want to extend the discussion of Sec.~\ref{steadystates} to dark states of networks with an arbitrary number of spins that coupled to one or many chiral waveguides. The analysis presented above for two and four spins will thereby serve as guide. In particular, we show that the conditions (i)-(iv) already anticipated in Sec.~\ref{dynamical} are sufficient for a system to have a unique pure dark state if $N$ is \emph{even}. In addition, we show that the structure of this steady state is in general of the form given in Eqs.~(\ref{General_SS}) and (\ref{multimer}), i.e.,~the system factorizes into a product of clusters, and we connect this structure to the properties of the bare spin system, in particular to the detuning pattern $\delta_j$.

\subsection{Cascaded channel and dimerization}\label{Proof_Cascaded}

We first take a detour [cf.~Sec.~\ref{Proof_Cascaded}] and consider not a chiral, but a \emph{cascaded} setup instead. The cascaded problem is simpler, inasmuch as the unidirectional flow of information allows an analytic solution from ``left to right.'' Using this property, it was shown in Ref.~\cite{{Stannigel:2012jk}} that the steady state of a cascaded spin system (under conditions specified below) has a unique pure steady state in which the system dimerizes; i.e.,~the steady state is of the form 
\begin{align}
\ket{\Psi}\!=\!\bigotimes_{j=1}^{N/2}\ket{D(\alpha_{2j-1})}_{2j-1,2j}.\label{dimerizedstatee}
\end{align}
Here each spin $j\!=\!1,\dots,N$, pairs up with one of its neighbors to form a dimer and decouples from the rest of the chain.

To this end, let us consider a system of $N$ spins that are coupled via a unidirectional channel as described by the cascaded ME \eqref{Cascaded_ME}. A defining property of Eq.\,\eqref{Cascaded_ME} is that information flows only in one way, specifically in the propagation direction of the photons along the unidirectional waveguide. While this is evident from the physical picture underlying the ME, one can also see this also on a formal level. For example, it is possible to calculate the equations of motion for the ``first'' or ``leftmost'' spin along the cascaded channel, by simply tracing out the degrees of freedom of all other spins in \,\eref{Cascaded_ME}, obtaining
\begin{align}\label{OBE_1spin}
\dot{\rho}_{1}=\!&-i[-\delta_{1} \sigma_1^{\dag}\sigma_1+\Omega_1(\sigma_1+\sigma_1\dg),\rho_{1}]+\gamma_R\mathcal{D}[\sigma_1]\rho_{1}.
\end{align}
The above ME (\ref{OBE_1spin}) is closed, meaning that the first spin is independent of the state of all other spins and reflecting the unidirectionality of the system. Note that \eref{OBE_1spin} is the well-known optical Bloch equation for a single driven two-level system and thus its steady state is in general mixed.

More interesting is the equation of motion for the density operator of the first two spins $\rho_{1,2}$, which is obtained from Eq.\,\eqref{Cascaded_ME} analogously to Eq.\,\eqref{OBE_1spin} and reads
\begin{align}\label{Cascaded_2spins}
\dot{\rho}_{1,2}=&-i \sum_{j=1,2}[-\delta_{j} \sigma_j^{\dag}\sigma_j\!+\!\Omega_j (\sigma_j\!+\!\sigma_j\dg),\rho_{1,2}]\nonumber\\&-
\frac{\gamma_R}{2}[\sigma_2^{\dag}\sigma_1\!-\!\sigma_1\dg \sigma_2,\rho_{1,2}]+\gamma_R\mathcal{D}(\sigma_1\!+\!\sigma_2)\rho_{1,2}.
\end{align}
Again, the equation of motion of the first two spins does not depend on the state of any other spin, since the first two spins do not notice the presence of the others in the cascaded setup. Importantly, Eq.~\eqref{Cascaded_2spins} is a special case (with $\Delta\gamma=\gamma_R$) of the chiral master equation for two spins, already analyzed in Sec.~\ref{two_spins}. There we showed that after a characteristic time $t_{D}$ [cf.~Eq.\,\eqref{Timescale_dimer}], the two spins dynamically purify into the dimer state $\ket{D(\alpha_{1})}$ in Eq.\,(\ref{Dimer}), provided $\delta_2=-\delta_1$ and $\Omega_2=\Omega_1=\Omega$. The corresponding singlet fraction is given as in Eq.\,(\ref{singletfraction}), but here with $\Delta\gamma=\gamma_R$. 

\begin{figure*}[t]
\includegraphics[width=1\textwidth]{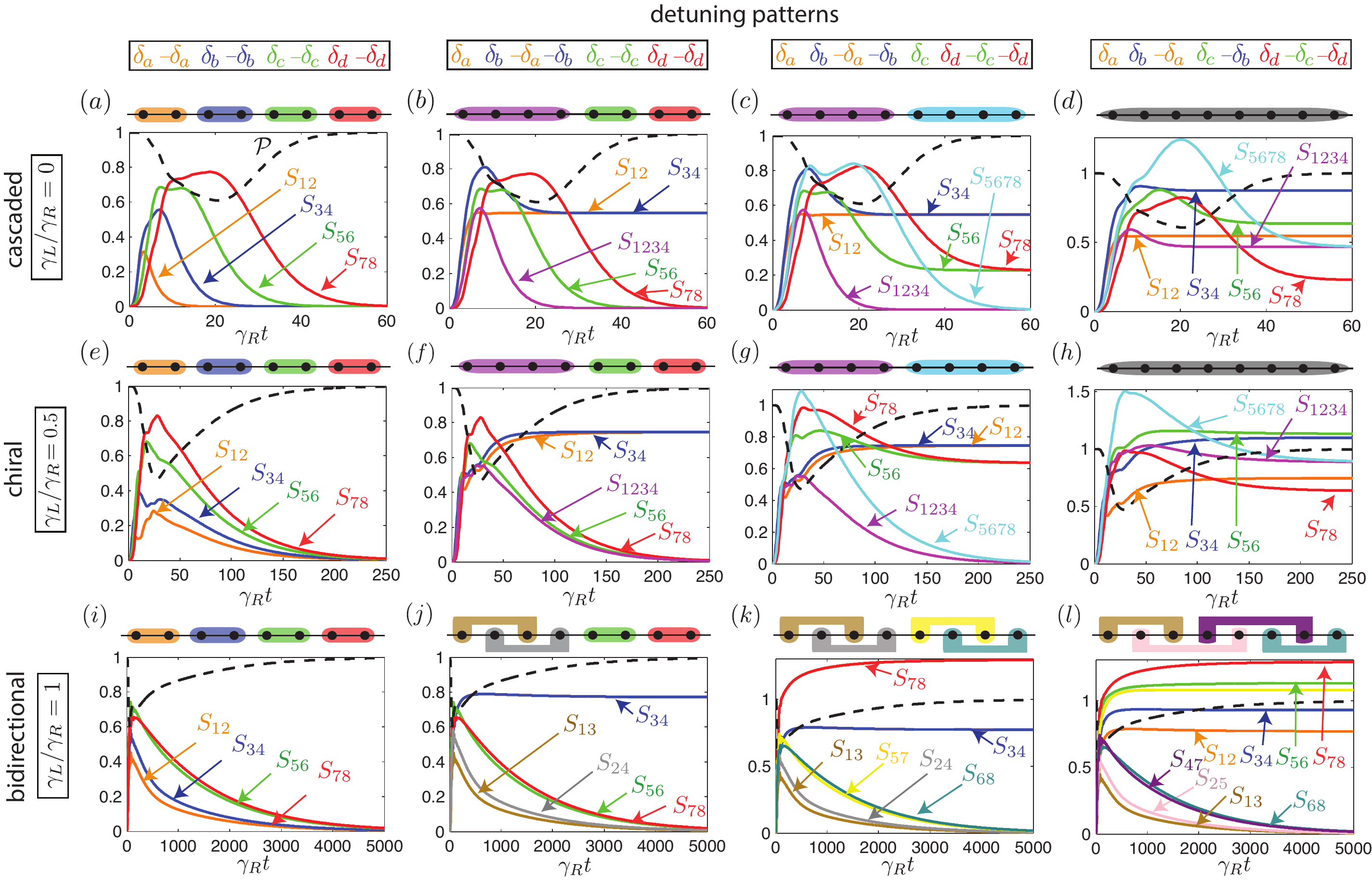}
\caption{(Color online) Dynamical purification in the cascaded setup ($\gamma_L\!=\!0$, first row), chiral setup ($\gamma_L\!=\!0.5\gamma_R$, second row), and bidirectional setup ($\gamma_L\!=\!\gamma_R$, third row). We show the entropies of reduced density matrices $S_{j_1,j_2,\dots}$ of spins $\{j_1,j_2,\dots\}$ (colored solid lines) and the purity of the total state $\mathcal{P}$ (dashed black lines) as a function of time. In the first column the detuning pattern is chosen such that the steady state dimerizes, which is signaled by a vanishing entropy of the reduced density matrix of the corresponding spin pairs (see text). While in the cascaded setup the system purifies from left to right, in the chiral case the system purifies as a whole. In the second column the detuning pattern is chosen such that the steady state breaks up into a tetramer and two dimers. The last two columns show analogous situations for detuning pattern corresponding to two tetramers and an octamer, respectively. Note that in the bidirectional case ($\gamma_L=\gamma_R$, last row), the steady state is always dimerized, but the dimers can be nonlocal, depending on the detuning pattern. Other parameters are $\Omega=0.5\gamma_R$, $\delta_a=0.6\gamma_R$, $\delta_b=0.4\gamma_R$, $\delta_c=0.2\gamma_R$ and $\delta_d=0.1\gamma_R$.}\label{fig:Purification}
\end{figure*}

Since this state is pure, the first two spins cannot be entangled with any of the other spins and thus the state of the total system for times $t\gg t_D$ has the form $\rho(t)=\ket{D}_{1,2}\!\bra{D}\otimes \rho_{3,\dots,N}(t)$. Once the first two spins are in the dark state $\ket{D}_{1,2}$, they no longer scatter photons and therefore do not affect the dynamics of any of the other spins. 
The equation of motion for $\rho_{3,4}$ then decouples not only from spins $\{5,\dots,N\}$ due to the cascaded nature of the problem, but also from the first pair forming the dark state. The ME for $\rho_{3,4}$ is therefore closed and given by an expression analogous to Eq.\,(\ref{Cascaded_2spins}). As for the first pair, also this second pair is driven into the pure dark state $\ket{D(\alpha_{3})}_{3,4}$ if $\delta_4=-\delta_3$ and $\Omega_4=\Omega_3=\Omega$. This argument can be iterated to show that the dimerized state  in \eref{dimerizedstatee} is the unique steady state of a cascaded spin chain with an even number of spins $N$, driven homogeneously $\Omega_j=\Omega$, and with a ``staggered'' detuning pattern $\delta_{2j}=-\delta_{2j-1}$, $j=1\dots N$. Remarkably this iterative purification from left to right is not only a mathematical trick to solve for a dark state, but it is also realized physically, meaning that the cascaded system is indeed dynamically purified from left to right, as shown in Fig.~\ref{fig:Purification}(a). There we numerically calculate the timeevolution of the entropies of spin pairs, $S_{i,j}\equiv-\textrm{Tr}\{\rho_{i,j}\ln\rho_{i,j}\}$, signaling the successive formation of dimers as $S_{2j-1,2j}\rightarrow 0$, for different pairs at different times. They are separated by the relaxation times $t_D$ given in \eref{Timescale_dimer} and the time scale to form the full dimerized state in the cascaded setup is proportional to the number of spins $t_{\rm ss}\sim Nt_D/2$ \cite{footnote_cascaded}.

If the total number of spins is odd, all spins except the last one are driven into such dimers. This last spin simply factorizes off and  goes to a mixed steady state, as its dynamics is described by a ME of the same form as Eq.~\eqref{OBE_1spin}, once all other spins reach the dimerized dark state [cf.~Fig.~\ref{odd}(a)].

\subsection{Dimerization in a chiral channel}\label{Proof_Chiral}

For spins coupled to a chiral channel ($0<\gamma_L<\gamma_R$), an iterative solution for the steady state as in the cascaded setup is not possible due to the non-unidirectional flow of information. However, we have already seen for $N=2$ and $N=4$ that the generic chiral ME \eqref{chiralME} also has dark steady states if the general conditions (i)-(iv) of Sec.\,\ref{dynamical} are satisfied. We show below that this holds true also for \emph{arbitrary even} $N$.

We start by showing that under the same conditions as in the cascaded case, also the chiral ME \eqref{chiralME} drives the spins into a dimerized steady state, which in Sec.~\ref{Proof_Chiral_Multipartite} is the starting point to obtain more complex multipartite entangled dark states. Since the solution of the cascaded ME relies heavily on its unidirectional character, it is quite remarkable that it is this solution that allows us to construct also the dark states of its chiral counterpart. In fact any dark steady state of the cascaded system can be obtained also in a chiral setup, as we show in the following.

First, we note that under condition (i) of Sec.\,\ref{dynamical} one finds the relations
\begin{align}\gamma_RH_L=-\gamma_LH_R,\quad c_L=c_R.\end{align} 
and thus the chiral ME (\ref{chiralME}) can be written as a sum of a cascaded Liouvillian, whose strength is replaced by $\Delta\gamma\geq 0$, and an additional Lindblad term with the single collective jump operator $c_R$ of strength $2\gamma_L$:
\begin{align}\label{chiral_likecascaded}
\dot\rho&\!=\!-\frac{i}{\hbar}\!\Big[H_{{\rm sys}}\!+\!\frac{\Delta\gamma}{\gamma_R} H_R,\rho\Big]\!\!+\!\!\frac{\Delta\gamma}{\gamma_R}\mathcal{D}[c_R]\rho\!+\!2\gamma_L\mathcal{D}[c_R]\rho.
\end{align}
As shown in the Sec.~\ref{Proof_Cascaded}, the dimerized state is the unique pure steady state of the cascaded part of Eq.~(\ref{chiral_likecascaded}). By construction this dark state is annihilated by the \emph{single} collective jump operator $c_R$, such that it is also a dark state of the additional term in the chiral setup. Thus, any unique pure steady state of the cascaded ME is also the steady state of the corresponding chiral ME, with the identification $\gamma_R\rightarrow \Delta\gamma$. Moreover, it is also guaranteed to be unique as long as $\Delta\gamma>0$. In particular, the dimerized state \eqref{dimerizedstatee} is also the steady state of the chiral ME, where only the singlet fraction \eqref{singletfraction} is renormalized with respect to the cascaded case. We note that this construction requires $\Delta\gamma\neq0$ and thus can not be extended naively to the bidirectional setting of Sec.~\ref{dickemodel}. This special case will be discussed in Sec.~\ref{Nnonlocalpairs}.

\begin{figure}[t]
\includegraphics[width=0.45\textwidth]{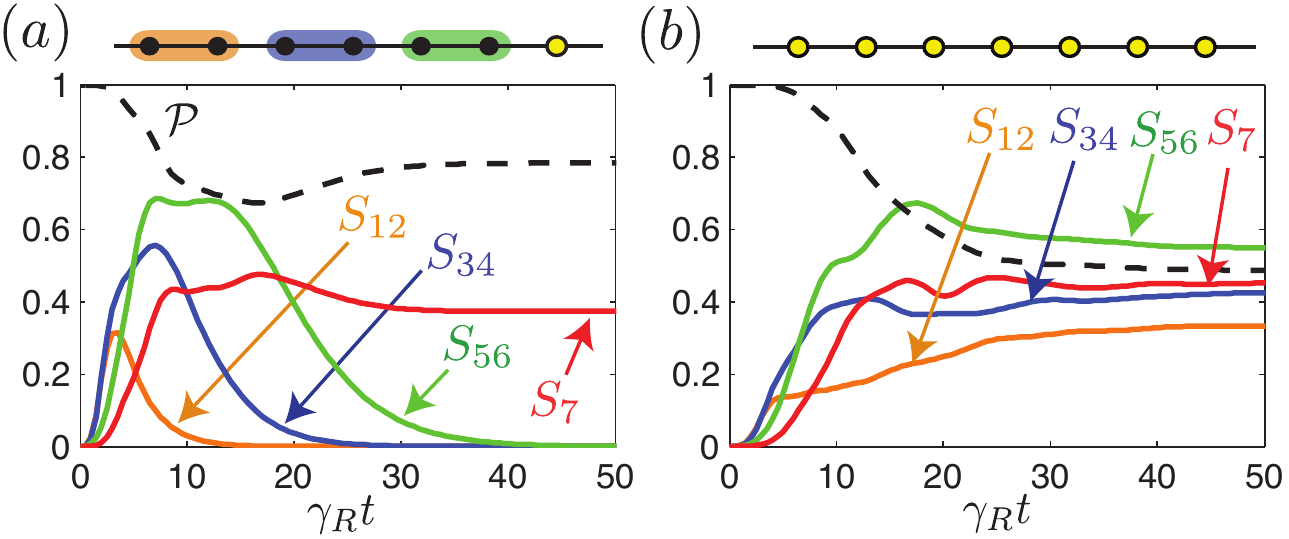}
\caption{(Color online) Typical behavior of a system with an odd number of spins in the case of $N=7$. 
We show the entropies of reduced density matrices $S_{j_1,j_2,\dots}$ of spins $\{j_1,j_2,\dots\}$ (colored solid lines) and the purity of the total state $\mathcal{P}$ (dashed black lines) as a function of time. 
(a) In the strict cascaded limit ($\gamma_L=0$), dimers are formed, but the last spin stays mixed and renders the steady-state non-dark (cf.~red and black dashed curves). (b) If $\gamma_L\neq 0$ the steady state is mixed, and no sub-structure is formed. Other parameters are the same as in Figs.~\ref{fig:Purification}(a) and \ref{fig:Purification}(b).}\label{odd}
\end{figure}

Note that the treatment of the cascaded case cannot be extended to the chiral one if the number of spins is \emph{odd}. In the cascaded setup, even though dimers are formed, the last unpaired spin scatters photons and thus the state is \emph{not dark} [cf.~Fig.~\ref{odd}(a)]. However, only \emph{dark} states of the cascaded ME are also steady in the chiral pendant. From a physical point of view this is clear, since the unpaired spin in the chiral setting will scatter photons to both sides of the chain and thus necessarily disturb any dimers that may have been formed between other spins, inhibiting a dimerization of the steady state [cf.~Fig.~\ref{odd}(b)].

Even though the cascaded and the chiral ME have the same steady states (for even $N$), the dynamics of the two systems in how this steady states is approached is rather different. While in the cascaded setup the spin chain purifies successively from left to right due to the unidirectional flow of excitations [cf.~Fig.~\ref{fig:Purification}(a)], in the chiral case the system purifies ``as a whole'' [cf.~Fig.~\ref{fig:Purification}(e)].

\subsection{Multipartite entanglement in a chiral spin chain}\label{Proof_Chiral_Multipartite}

The above discussion shows that the spin chain is driven into a pure dimerized steady state if driven homogeneously and with a ``staggered'' detuning pattern $\delta_j$ such that $\delta_{2j}=-\delta_{2j-1}$. For $N=4$ we found in Sec.~\ref{tetramersubsec} that also for \emph{permutations} of this detuning pattern the system has dark states, which are no longer dimerized, but rather four-partite entangled. It turns out that this concept can be generalized to any even number of spins $N$. In fact, 
a chiral spin chain driven with a detuning pattern obtained by any permutation $p\in S_N$ of the staggered one, i.e.,
\begin{align}\label{permuteddetuning}
\delta_{p(2j)}=-\delta_{p(2j-1)},
\end{align} 
goes into a pure steady state. Moreover, this steady state can be multipartite entangled. For the cascaded case this was shown in Ref.~\cite{Stannigel:2012jk} and - as for the dimerized state - the solution carries over also to the chiral setting.

\begin{figure}[b]
\includegraphics[width=0.45\textwidth]{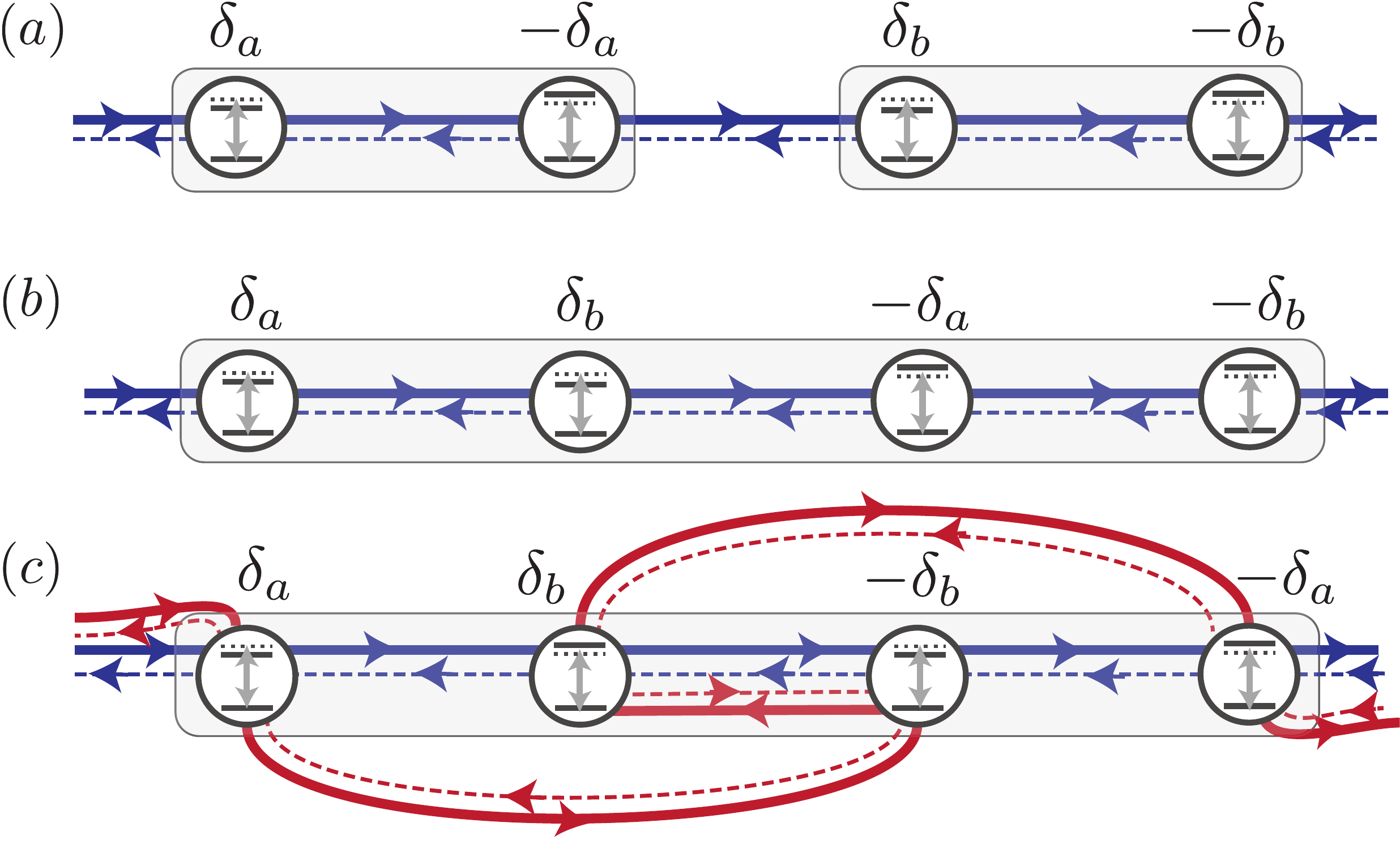}
\caption{(Color online) (a) Dimerized state as the steady state of the chiral spin chain, when driven on resonance with a staggered detuning pattern. (b) Tetramerized steady state, when the spin chain is driven with a permuted detuning pattern with respect to (a). (c) Simple 2-waveguide chiral network with a tetramerized pure steady state. The connection between the states generated in these 3 different situations is outlined in Secs.~\ref{Proof_Chiral}-\ref{multiwaveguideN}. 
}\label{fig:unitaryconnection}
\end{figure}

Specifically, there is a unitary mapping $U(p)$ that leaves the chiral master equation form-invariant up to permutations $p$ of the detunings, and therefore the corresponding steady states are connected by this transformation. Starting from the dimerized state $\ket{\Psi}$ one can thus construct dark states $U(p)\ket{\Psi}$, corresponding to MEs with more complex detuning patterns. To construct $U(p)$, we first consider the unitary transformation 
\begin{align}\label{Transposition}
\mathcal{U}_{j}(\vartheta)=\exp\lr{i\frac{\vartheta}{2}\vec{\sigma}_{j}\cdot\vec{\sigma}_{j+1}},
\end{align}
acting on two neighboring spins $j$ and $j+1$, where $\vec{\sigma}_j\equiv\big(\sigma_j^{x},\sigma_j^{y},\sigma_j^{z}\big)$ is the vector of Pauli matrices for spin $j$. One finds that for $\vartheta=\arctan\big\{(\delta _{j+1}-\delta_j)/\Delta\gamma \big\}$ the chiral ME \eqref{Dickebroken} is invariant up to the swap of the detunings $\delta_j\leftrightarrow \delta_{j+1}$ (cf.~ Ref.~\cite{Stannigel:2012jk}). Therefore, on the level of the steady states, interchanging the detunings between two neighboring spins  corresponds to applying the entangling operation in Eq.~\eqref{Transposition}, on the involved subsystems. For instance, the detuning patterns in Figs.~\ref{fig:unitaryconnection}(a) and \ref{fig:unitaryconnection}(b) differ by the exchange of $\delta_2\leftrightarrow\delta_3$. This is reflected in the structure of the corresponding steady state, inasmuch as in Fig.~\ref{fig:unitaryconnection}(a) it is dimerized, while in Fig.~\ref{fig:unitaryconnection}(b) it forms a tetramer. 

Since any permutation $p$ can be decomposed into a sequence of such pairwise transpositions, the  corresponding unitary $U(p)$ is given by a product of pairwise transformations of the form \eqref{Transposition}, and thus the structure of the steady state can be easily understood by constructing it via a sequential application of these entangling gates starting from the dimerized state. For example, we show in Figs.~\ref{fig:Purification}(a)-\ref{fig:Purification}(h) the timeevolution towards different types of ``clusterized'' steady states in systems of $N\!=\!8$ spins, including dimerized states [cf.~Figs.~\ref{fig:Purification}(a) and \ref{fig:Purification}(e)], tetramerized states [cf.~Fig.~\ref{fig:Purification}(c) and \ref{fig:Purification}(g)], octamers [cf.~Figs.~\ref{fig:Purification}(d) and (h)], but also heterogeneous cluster sizes [cf.~Figs.~\ref{fig:Purification}(b) and (f)]. As already discussed in the example of the dimerized states above, also in this general setting there is a difference between the cascaded and the chiral setup, inasmuch the unidirectional character of the cascaded setting is reflected in the order at which the clusters purify.

\subsection{Many Chiral Waveguides}\label{multiwaveguideN}

A single chiral waveguide breaks the left-right symmetry and therefore introduces an ordering of the spins along it. When more waveguides are involved, as depicted for example in Fig.~\ref{fig:model}(b), the situation becomes more complex, since the order of the spins along each of them can differ. Remarkably, in this more general context, pure steady states are still possible. A complete characterization of the possible dark states in all chiral networks described by the ME \eqref{MultiME} is beyond the scope of this paper, but we rather want to show this for some simple cases. For instance, we are interested in the situation where all $M$ waveguides couple to all $N$ spins exactly once and where conditions analogous to (i)-(iii) of Sec.~\ref{dynamical} are satisfied for each waveguide. In particular, condition (i) simplifies the discussion drastically since the jump operators corresponding to emission of photons at both outputs of all waveguides are then the same, and equal to the collective jump operator $c=\sum_j \sigma_j$ discussed earlier.

The simplest nontrivial such network consists of two chiral waveguides, $m=1,2$ \big(with decay asymmetry $\Delta\gamma^{(m)}\equiv\gamma_R^{(m)}-\gamma_L^{(m)}>0$\big), where the order of two neighboring spins along the first and second waveguide is interchanged. Such a system is depicted in \Fref{fig:unitaryconnection}(c). 
It turns out that the corresponding ME can be unitarily mapped to the one of a set of spins coupled to a \emph{single} chiral waveguide \eqref{Dickebroken}, with different detunings. 
In fact one can show that this is achieved via the unitary given in \eref{Transposition} with a choice of $\vartheta$ such that 
\begin{align}
\tan(\vartheta)=\frac{\delta_j\!-\!\delta_{j+1}\!\pm\!\sqrt{(\delta_j-\delta_{j+1})^2\!+\!4\Delta\gamma^{(1)}\Delta\gamma^{(2)}}}{2\Delta\gamma^{(1)}}.
\end{align}
Under this transformation, this two-waveguide network maps onto a single-waveguide one with $\gamma_L=\gamma_L^{(1)}+\gamma_L^{(2)}$ and $\gamma_R=\gamma_R^{(1)}+\gamma_R^{(2)}$. 
Moreover, the detunings of spins $j$ and $j+1$ transform as
$\delta_j\rightarrow (\delta_j+\delta_{j+1})/2+\varepsilon/2$ and $\delta_{j+1}\rightarrow (\delta_j+\delta_{j+1})/2-\varepsilon/2$, with 
\begin{align}
\varepsilon&\equiv\!(\Delta\gamma^{(1)}\!+\!\Delta\gamma^{(2)})\sin(2\vartheta)\!+\!(\delta_j-\delta_{j+1})\cos(2\vartheta),
\end{align}
whereas all others are left invariant. From the discussion in Sec.~\ref{Proof_Chiral_Multipartite} we can thus infer that the steady state is pure if this transformed pattern satisfies condition (iv) of Sec.~\ref{dynamical}. For example, the situation depicted in Fig.~\ref{fig:unitaryconnection}(c) can be mapped into a single chiral waveguide similar to Fig.~\ref{fig:unitaryconnection}(b) with a detuning pattern $\delta_j\!=\!\{\delta_a,\varepsilon/2,-\varepsilon/2,-\delta_a\}$ and thus has a pure steady state.

\begin{figure}[t]
\includegraphics[width=0.49\textwidth]{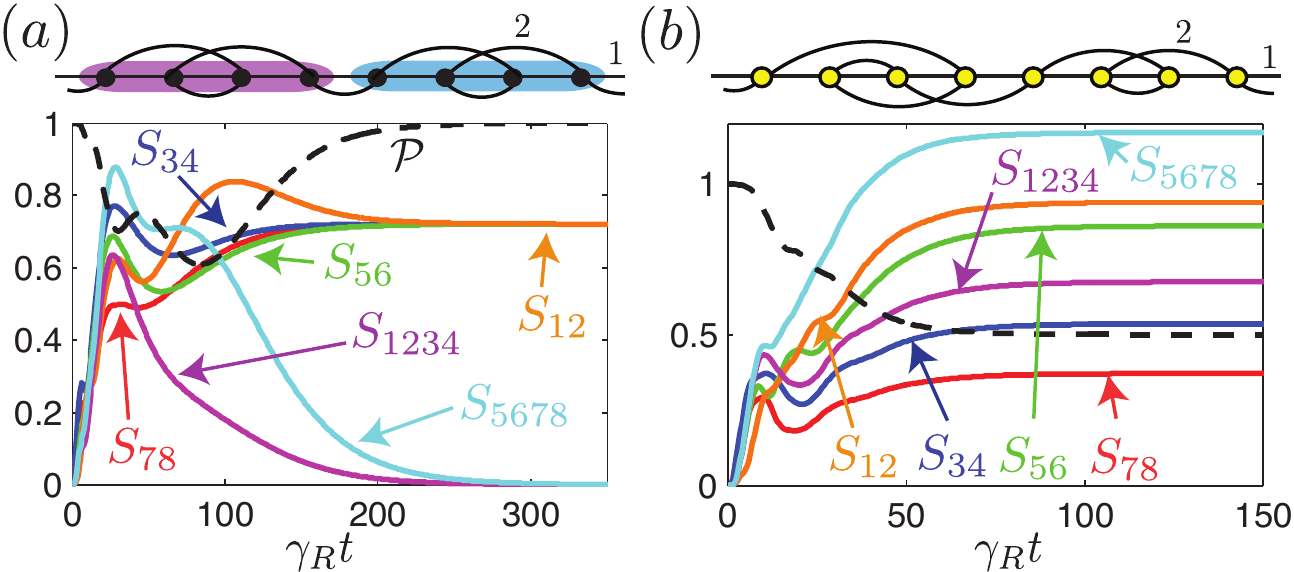}
\caption{(Color online) Steady states in multiple-waveguide chiral networks. We show the entropies of reduced density matrices $S_{j_1,j_2,\dots}$ of spins $\{j_1,j_2,\dots\}$ (colored solid lines) and the purity of the total state $\mathcal{P}$ (dashed black lines) as a function of time. (a) Pure tetramerized state as dark state of a 2-waveguide network. (b) 2-waveguide network with a wiring such that the steady state is mixed and without internal structure. 
Parameters as in Fig.~\ref{fig:purification_teaser}(c).}
\label{fig:Multi}
\end{figure}

This construction can be iterated to interchange the order of more spins along the waveguides and also to introduce more chiral waveguides. For example in Fig.~\ref{fig:Multi}(a) we show how a two-waveguide network can be wired to lead to a pure tetramerized  steady state. However, not every multiple-waveguide chiral network can be mapped in this manner to a single-waveguide chiral network [cf.~Fig.~\ref{fig:Multi}(b) for an example] and thus a completely general setting needs a different approach, which is beyond the scope of this paper.

\subsection{Special Case: Bidirectional channel}\label{Nnonlocalpairs}

The case $\gamma_L=\gamma_R$ has to be treated separately, since a unique steady state can only form if the permutation symmetry between all spins is broken. However in the absence of chirality $\Delta\gamma=0$, this is not guaranteed. In particular if some detunings are equal the symmetry is partially restored, leading to nonunique steady states, depending on the initial conditions. On the other hand, if all detunigs are different, the permutation symmetry is again fully broken, and the steady state is unique also in the bidirectional case. Additionally, if conditions (i)-(iv) of Sec.\ref{dynamical} are fulfilled, the steady state is dark and -in contrast to the chiral case-  always dimerizes.  Spins with opposite detuning pair up in dimers and factorize off from the rest of the system, even if they are not nearest neighbors. Unlike the chiral setting, these dimers can not be entangled by interchanging the detunings of different spins. This is related to the fact that the unitary $U$ corresponding to such a swap [cf.~\eref{Transposition}] is not an entangling gate for $\Delta\gamma=0$. This behavior is illustrated in the last row of Fig.~\ref{fig:Purification}. In the absence of chirality, the coupling between subspaces of different permeation symmetry is weaker, and correspondingly the timescale to approach this steady state is longer than in the chiral or cascaded counterparts (cf.~Fig.~\ref{fig:Purification}).

\subsection{Remarks on less restrictive assumptions for dark states}

We remark that both conditions (i) and (ii) stated in Sec.~\ref{dynamical} can be trivially relaxed in some situations. In particular, for a clusterized state of the form in \eref{General_SS} (with $N_m\geq 2$), these conditions need to be fulfilled only within each cluster of spins. For example, dark states still form if the coherent driving field $\Omega_j$ varies from cluster to cluster. Similarly, the spacings of the spins has to be commensurate with the photon wavelength only for spins within each cluster. This simply reflects the fact that each cluster can be dark independently, since in that case it does not emit any photons into the waveguide and thus completely decouples from all other spins. 

\subsection{Imperfections}

\begin{figure}[t]
\includegraphics[width=0.49\textwidth]{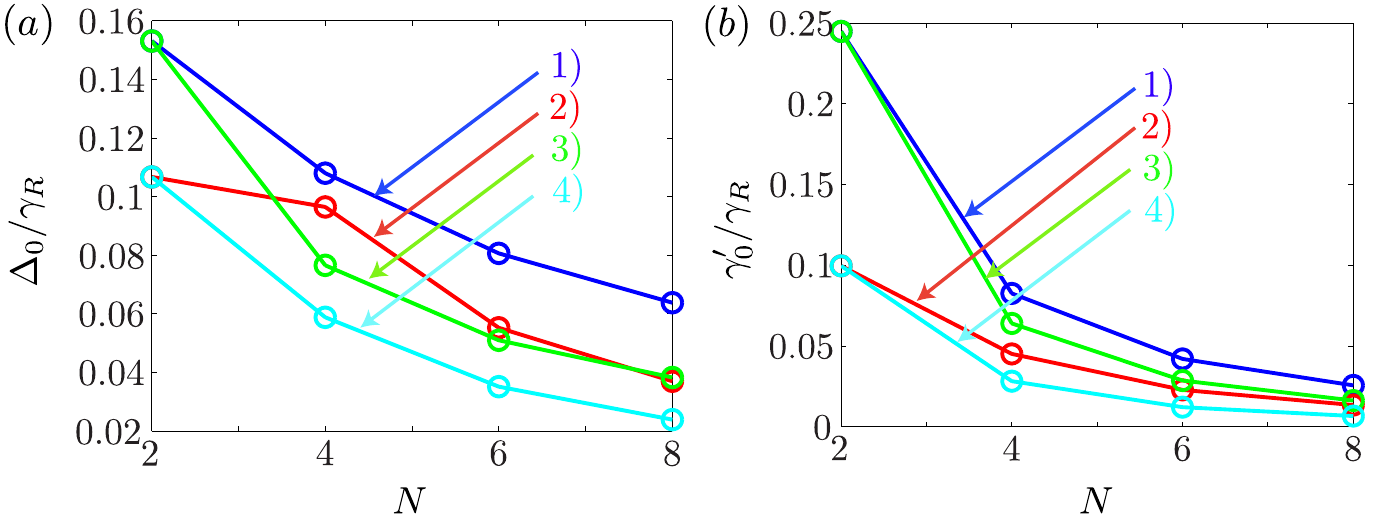}
\caption{(Color online) Imperfections for different system sizes. We consider a homogeneous offset in the detuning $\Delta$ on top of the ideal detuning pattern in (a) and additional on-site decay channel with decay rate $\gamma'$ in (b). For small $\Delta$ the purity of the steady states behaves like $\mathcal{P}=1-(1/2)(\Delta/\Delta_0)^2$, whereas for on-site decay, it scales linearly as $\mathcal{P}=1-\gamma'/\gamma'_0$ (see text). The figure shows the corresponding error susceptibilities $\Delta_0$ in (a) and $\gamma'_0$ in (b) for systems with $N=2,4,6,8$ spins. Parameters: (1) and (2) show the fully $N$-partite entangled situation with an ideal detuning pattern satisfying $\delta_1=\delta_N=0$ and $\delta_{2j}\!=\!-\delta_{2j+1}\!=\!0.3\gamma_R$, else. (3) and (4) show the dimerized situation $\delta_j=0$. For the decay asymmetries we choose $\gamma_L/\gamma_R=0$ in (1) and (3), $\gamma_L/\gamma_R=0.3$ in (2) and (4). We further fix $\Omega/\gamma_R=0.5$.}
\label{fig:error2468}
\end{figure}

An important question, so far discussed only in the example of $N\!=\!2$ spins in Sec.~\ref{two_spins}, is the error susceptibility of the steady state against various types of imperfections. In Fig.~\ref{fig:error2468} we numerically calculate the error robustness for different kinds of setups as a function of the size of the spin chain $N$. In particular, in Fig.~\ref{fig:error2468}(a) we show the effect of a homogeneous offset $\Delta$ on top of detuning patterns that would be consistent with pure steady states. For a finite $\Delta$ the steady state is no longer pure and its purity decreases as $\mathcal{P}=1-(1/2)(\Delta/\Delta_0)^2+\mathcal{O}(\Delta^4)$, such that the the error susceptibility is quantified by $\Delta_0$. This type of error is only of  second order in $\Delta$, since $\mathcal{P}(\Delta)=\mathcal{P}(-\Delta)$, which can be shown by noting that the ME corresponding to $\Delta$ and the one corresponding to $-\Delta$ can be unitarily mapped into each other. One can very clearly see that the error susceptibility increases with system size, and moreover that the chiral setting is more vulnerable than the cascaded counterpart. This can be understood intuitively, since any imperfection will disturb the formation of the dark state, e.g.,~dimers. Moreover, an imperfectly formed dimer scatters photons affecting also the other parts of the system. While all pairs can be disturbed by such photons in the chiral setting, in the cascaded setting they act as an additional perturbation only on pairs on its left.

In Fig.~\ref{fig:error2468}(b) we show the effect of a finite on-site decay of each spin with a rate $\gamma'$. In this case the purity depends linearly on $\gamma'$; i.e., $\mathcal{P}=1-\gamma'/\gamma'_0+\mathcal{O}(\gamma'^2)$. Therefore, the quantity $\gamma'_0$ shown in Fig.~\ref{fig:error2468}(b) is a rough bound on the maximum decay rate $\gamma'$ still allowed to see the effect of a dynamical purification. One finds, again, that chiral systems are more error prone than their cascaded counterpart and that the control of the on-site decay $\gamma'$ becomes more crucial for larger systems. While such imperfections are in general hard to control in current photonics realizations of the spin network \cite{Reitz:2013bs,Goban:2013wp}, this would be intrinsically absent in a cold-atom realization \cite{Ramos:2014ut}.

\subsection{Quantum Trajectories calculations}

\begin{figure}[t]
\includegraphics[width=0.50\textwidth]{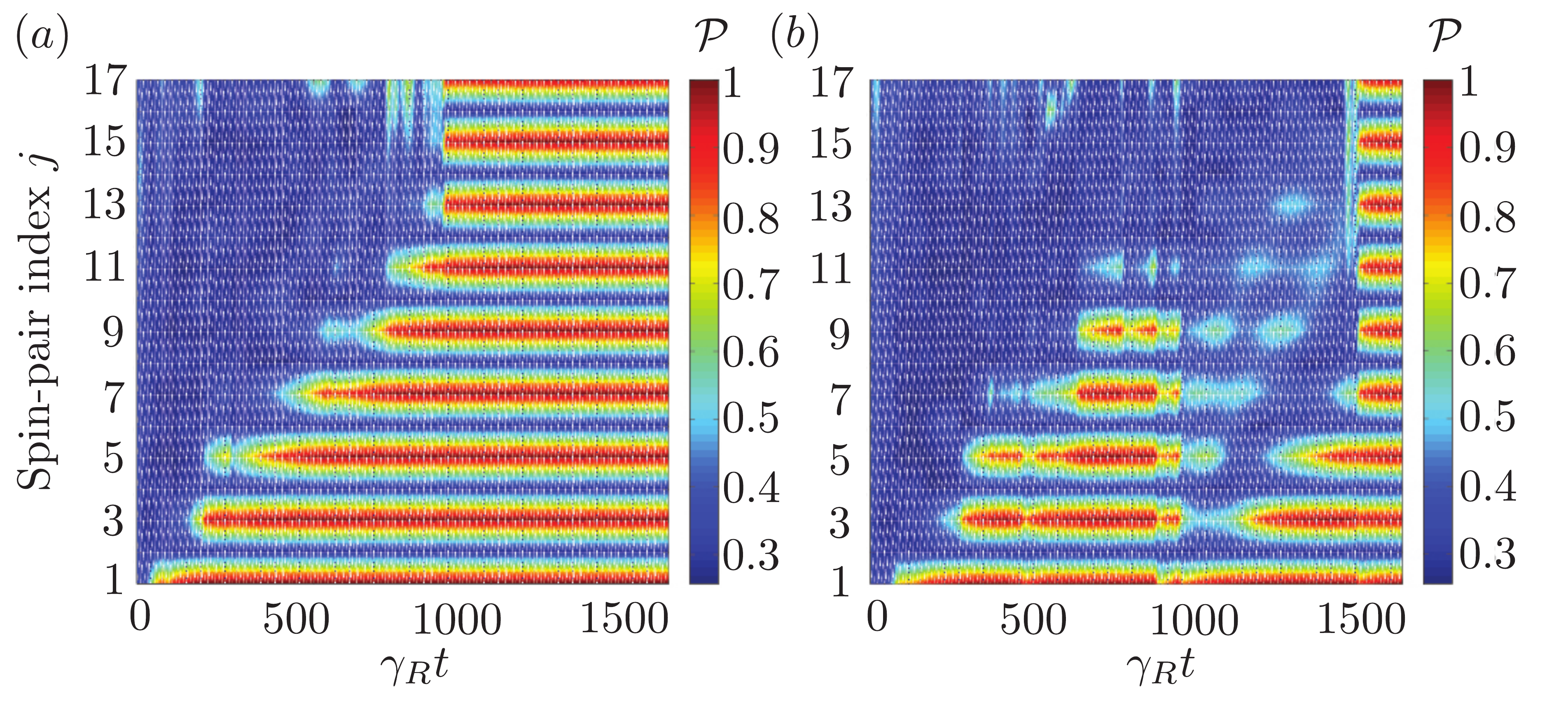}
\caption{(Color online) (a) Illustration of the time-dependent process of formation of dimers, through a single random trajectory in a quantum trajectories calculation, as described in the text, with $\gamma_L=0$. We plot the purity of each spin pair $j$, determined by $\mathcal{P}_{j,j+1}={\rm Tr}\{\rho_{j,j+1}^2\}$, where $\rho_{j,j+1}$ is the reduced density operator for spins $j$ and $j+1$, shown as a function of the spin pair index $j$ and time $t$. Here, we take a chain of $N=18$ spins, and choose $\Omega=1.8\gamma_R$. The shading is interpolated across the plot, so that we clearly see the formation of pure spin pairs between all pairs $(j,j+1)$ with $j$ odd, while the reduced density operators for all pairs $(j,j+1)$ with $j$ even remain in a mixed state. (b) Same as in (a) but with $\gamma_L=0.05\gamma_R$. We can clearly see that quantum jumps with $\gamma_L \neq 0$ can lead to breakup of already formed dimers, that tends to lengthen the process of reaching the steady state.} \label{fig:examplegl0}
\end{figure}

We can extend our calculations to larger spin chains, for example by integrating the ME \eqref{Cascaded_ME} using quantum trajectories methods \cite{Plenio:1998jk,Daley:2014hc}. In addition to averaging over trajectories to reproduce expectation values associated with the ME, these methods also give us an interpretation of the dynamics that would occur under continuous measurement of whether collective decay into the waveguide had occurred as a function of time \cite{QuantumNoise}. In Fig.~\ref{fig:examplegl0} we show example trajectories obtained by propagating an initial state with all of the spins in the ground state. Following the usual prescription, the states are propagated under the effective, non-Hermitian Hamiltonian, e.g.,~$H_{\rm sys}+H_{\rm eff}$ in the cascaded case, with collective jumps under the jump operator $c$ occurring in appropriately statistically weighted time steps \cite{Plenio:1998jk,Daley:2014hc}.

In Fig.~\ref{fig:examplegl0}(a) we show an example trajectory for the cascaded case $\gamma_L=0$, and we clearly see that the purity of the reduced state of each pair of spins increases in succession as the cascaded system evolves towards the steady state. In this plot, we also see how in the cascaded case the time scale for formation of the pairs increases linearly with the length of the chain. The dynamics of the chiral case with $\gamma_L\!=\!0.05\gamma_R$ is more complicated, as illustrated by the equivalent example trajectory in Fig.~\ref{fig:examplegl0}(b). With coupling in two directions, jumps can lead to a sudden decrease in the purity of a range of different spins, which then reestablish their purity in the subsequent time evolution. This type of process substantially slows the dynamics as $\gamma_L$ is increased, as predicted also for two spins in Eq.\,(\ref{Timescale_dimer}).

\subsection{Adiabatic preparation of the dark state}

\begin{figure}[t]
\includegraphics[width=0.5\textwidth]{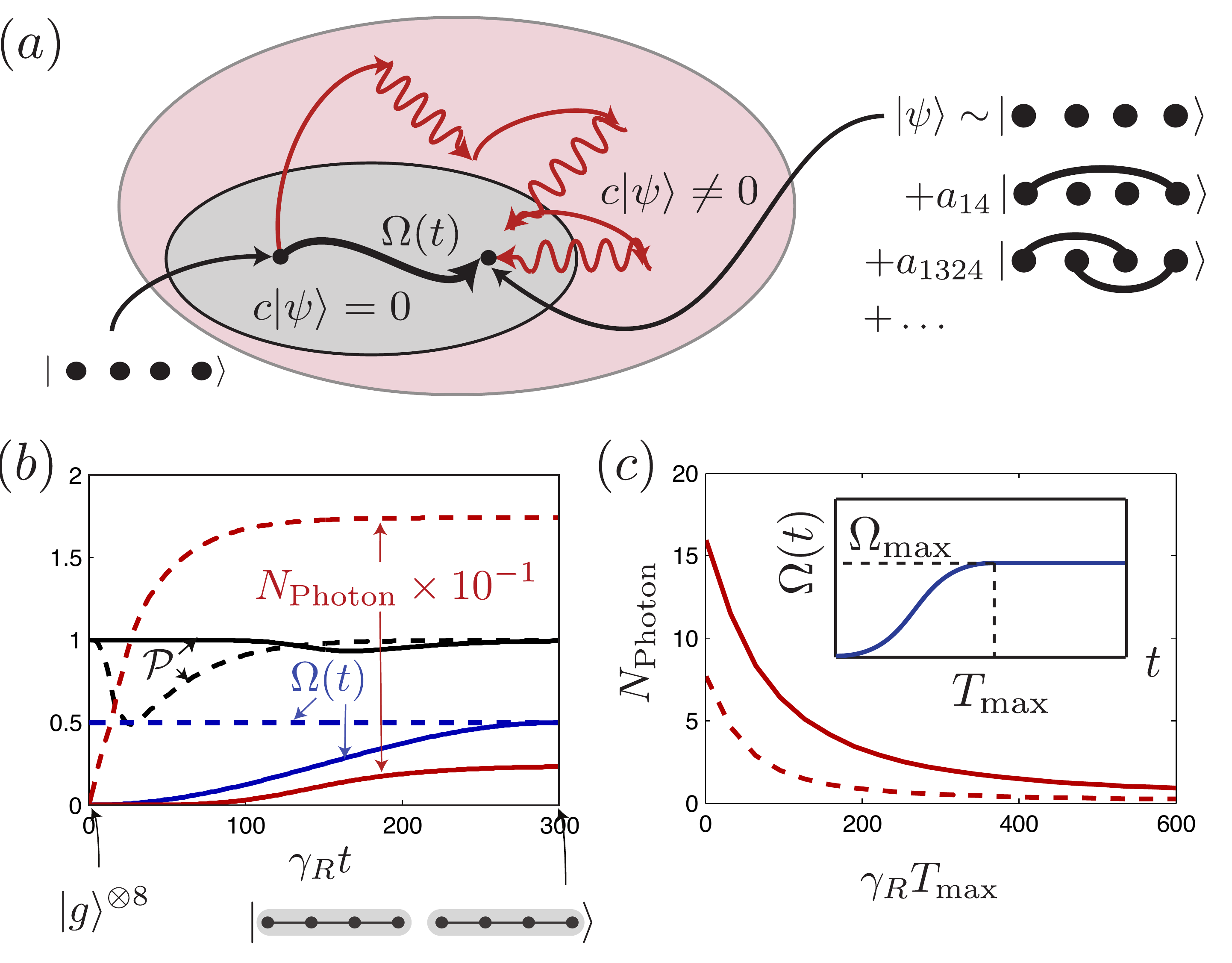}
\caption{(Color online) (a) Schematic illustration of different ways to ``cool'' to the many-body dark steady state (see text). The thick black arrow corresponds to an adiabatic path, while the red arrows indicate a nonadiabatic one. (b) Dynamical purification of a chain of $N\!=\!8$ spins into two tetramers initially in the state $\ket{g}^{\otimes N}$ for a sudden switch on of the constant coherent driving field (dashed lines), and for an ``adiabatic'' switching on of the driving field (solid lines). The black lines correspond to the purities $\mathcal{P}$ of the total system in the two cases. The total number of photons leaving the system in both cases is plotted in red; residual photons in the ``adiabatic case'' are due to non-adabatic effects stemming from a finite $T_{\rm max}\gamma_R=300$. Parameters are $\gamma_L/\gamma_R=0.5$, and the detuning pattern chosen is $\delta_j/\gamma_R=\{0,0.4, -0.4, 0, 0, 0.4, -0.4, 0\}$. (c) Total number of photons scattered for $N=6$ (solid line)  and $N=4$ (dashed line) as a function of the ramp-time $T_{\rm max}$. Parameters are $\Omega_{\rm max}=0.5\gamma_R$, $\gamma_L/\gamma_R=0.5$, $\delta_j/\gamma_R=\{0,0.4, -0.4, 0, 0, 0\}$ (solid line), and $\delta_j/\gamma_R=\{0,0.4, -0.4, 0\}$ (dashed line).}
 \label{fig:adiabatic}
\end{figure}

As a last side remark in this section, we note that the steady states discussed here can be reached dynamically in different ways. So far we considered the situation where $\Omega$ is constant in time, such that, starting with an initial state, e.g.,~$\ket{g}^{\otimes N}$, the driven system will scatter photons that leave the chiral waveguide at one of the two output ports until the dark state is reached. This scenario is the many-body analog of optical pumping in quantum optics \cite{HAPPER:1972fi}. Alternatively, one can reach the steady state without scattering a single photon by changing the coherent drive time-dependently $\Omega=\Omega(t)$, and in particular turning it on slowly. Then, the system follows adiabatically the instantaneous dark states corresponding to $\Omega(t)$, never leaving the nonemitting subspace defined by $c\ket{\psi}\!=\!0$ [cf.~Fig.~\ref{fig:adiabatic}(a)], reminiscent of the stimulated Raman adiabatic passage ($\textrm{STIRAP}$) in quantum optics \cite{Bergmann:1998ge}. In Fig.~\ref{fig:adiabatic}(b) we illustrate this on the example of a spin chain (initially in the trivial state $\ket{g}^{\otimes N}$) that reaches a teteramerized steady state. For $\Omega(t)=\Omega_{\rm max}$ the state purity initially decreases as the system scatters photons before it eventually purifies again into the entangled steady state. For an ``adiabatic'' switching on of the driving field according to $\Omega(t)=\Omega_{\rm max}\sin^2(\frac{\pi}{2}\frac{t}{T_{\rm max}})$ the system is almost pure, indicating that it follows the instantaneous dark state from the trivial initial state to the highly entangled final state. Figure~\ref{fig:adiabatic}(c) shows the total number of scattered photons $N_{\rm Photon}\equiv(\gamma_L+\gamma_R)\int_0^{t\rightarrow \infty}d\tau \textrm{Tr}\{c\dg c\rho(\tau)\}$ before the system relaxes to the steady state as a function of the turn-on time $T_{\rm max}$. One can clearly see that the total number of photons leaving the waveguide goes to zero with $T_{\rm max}\rightarrow \infty$. 

\section{Multipartite Entanglement detection via Fisher Information}\label{FisherSec}

In this section we discuss the possibility to detect the entanglement generated in the chiral spin networks discussed in this work. In particular, we are interested in the possibility to witness entanglement via the Fisher information and analyze its suitability in the present context.

Since the steady state is in general not only pure, but also
fragments into a product of multimers, state tomography can be efficient. For periodic detuning patterns the experimental cost for such a tomography does not scale exponentially with the system size, but linearly. Even though such a state tomography may be efficient, it still may be challenging to perform since it requires local measurements. 

An alternative route to analyze the entanglement properties is to use entanglement witnesses. The advantage of such witnesses is that they do not require full knowledge of the state, and therefore can be determined with a smaller set of measurements, that are potentially much simpler to perform as compared to state tomography. 
Recently, it has been shown in Refs.~\cite{Hyllus:2012ha} and \cite{Toth:2012hk} that the Fisher information can be used to witness multipartite entanglement. Moreover, the Fisher information has been measured in an experiment with cold atoms and used to detect entanglement \cite{Strobel:2014eg}. In the remainder of this section we review some properties of the Fisher information, its relation to entanglement and analyze up to what extent it can be used to detect the entanglement generated in the steady state of a chiral spin network.

\subsection{Fisher information and entanglement}

Originally, the Fisher information was introduced in the context of parameter estimation \cite{Braunstein:1994jl}. There, one is interested in distinguishing the state $\rho$ from the state $\rho_\theta=e^{-iG\theta}\rho e^{iG\theta}$, obtained by applying a unitary induced by a Hermitian generator $G$. To infer the value of $\theta$ one performs a measurement $M=\{M_\mu\}$, which in the most general case is given by a positive operator valued measure (POVM). The Fisher information $F[\rho,G,M]$ quantifies the sensitivity of this measurement and gives a bound on the accuracy to determine $\theta$ as
$(\Delta\theta)^2\geq 1/F$. In particular, the Fisher information is defined as \cite{Braunstein:1994jl}
\begin{align}
F[\rho,G,M]\equiv\sum_\mu \frac{1}{P(\mu|\theta)}\lr{\frac{\partial P(\mu|\theta)}{\partial\theta}}^2,\label{fisherdef}
\end{align}
where $P(\mu|\theta)\equiv\textrm{Tr}\{\rho(\theta)M_{\mu}\}$ is the probability to obtain the measurement outcome $\mu$ in a measurement of $M$ given the state $\rho(\theta)$. 

The Fisher information for an optimal measurement, i.e.,~the one that gives the best resolution to determine $\theta$, is called quantum Fisher information, and is defined as  $F_Q[\rho,G]\equiv \max_M F[\rho,G,M]$. In pure states it takes the simple form \cite{Braunstein:1994jl}
\begin{align}
F_Q[\ket{\psi}\bra{\psi},G]=4(\Delta G)^2,\label{quantumFisherInfo}
\end{align}
relating the quantum Fisher information to the variance of the generator $(\Delta G)^2\equiv\bra{\psi}G^2\ket{\psi}-\bra{\psi}G\ket{\psi}^2$.

There is an interesting link between quantum metrology and entanglement, inasmuch as entangled states can be useful to improve measurement sensitivities \cite{Pezze:2009ce,Toth:2014ew}. In particular, in Refs.~\cite{Hyllus:2012ha} and \cite{Toth:2012hk} it has been shown that the quantum Fisher information witnesses multipartite entanglement in spin systems, as the ones considered here. For linear generators $G=(1/2)\sum_{j=1}^N\vec{n}_j\cdot\vec{\sigma}_j$ (with with $|\vec{n}_j|=1$), the quantum Fisher information of a $k$-producible state, is bounded by \cite{Toth:2012hk,Hyllus:2012ha}
\begin{align}\label{Fisher_witness}
F_Q[\rho,G] \leq f(k,N)\equiv n k^2+(N-nk)^2,
\end{align}
where $n$ is the integer part of $N/k$. Therefore, a quantum Fisher information $F_Q[\rho,G]>f(k,N)$ witnesses $(k+1)$-partite entanglement. Notice that this criterion also applies for the Fisher information corresponding to any measurement $M$, since  $F_Q\geq F$. 
To witness entanglement via  \eref{Fisher_witness} it is desirable to use a generator $G$ that maximizes the quantum Fisher information. However, the optimal local rotation axes $\vec{n}_j$, corresponding to this generator, are in general dependent on the state and need to be determined numerically. For pure states it was shown in Ref.~\cite{Hyllus:2010ix} that the optimal quantum Fisher information $F_Q^{\rm max}[\ket{\psi}\bra{\psi}]\equiv\max_G F_Q[\ket{\psi}\bra{\psi},G]$ is given by 
\begin{align}
F_Q^{\rm max}=&\max_{\{\vec{n}_j\}} \sum_{i,j=1}^N\sum_{a,b=x,y,z}n_{i}^a\Gamma_{i,j}^{a,b}n_{j}^b,\quad \textrm{with}\label{Max_QFI}\\
\Gamma_{i,j}^{a,b}\equiv&\frac{1}{2}\bra{\psi}(\sigma_{i}^{a}\sigma_{j}^{b}+\sigma_{j}^{b}\sigma_{i}^{a})\ket{\psi}-\bra{\psi}\sigma_{i}^{a}\ket{\psi}\bra{\psi}\sigma_{j}^{b}\ket{\psi}.
\end{align}
Here $\sigma_{i}^{a}$ denotes the $a$th Pauli matrix on site $i$, and analogously $n_{i}^{a}$ denotes the $a$ component of $\vec{n}_i$. Thus, given the two-spin correlation function of a pure state, one has to solve a quadratically constrained quadratic problem. For a positive semidefinite $\Gamma_{i,j}^{a,b}$, efficient numerical algorithms (e.g.,~ semidefinite programming) are known \cite{Vandenberghe:1996te}. Moreover, an upper bound can be easily found in terms of the largest eigenvalue $\lambda_{\rm max}$ of $\Gamma_{i,j}^{a,b}$ and is given by $F_Q^{\rm max}\leq N\lambda_{\rm max}$ \cite{Hyllus:2010ix}. 

\begin{figure}[t]
\includegraphics[width=0.5\textwidth]{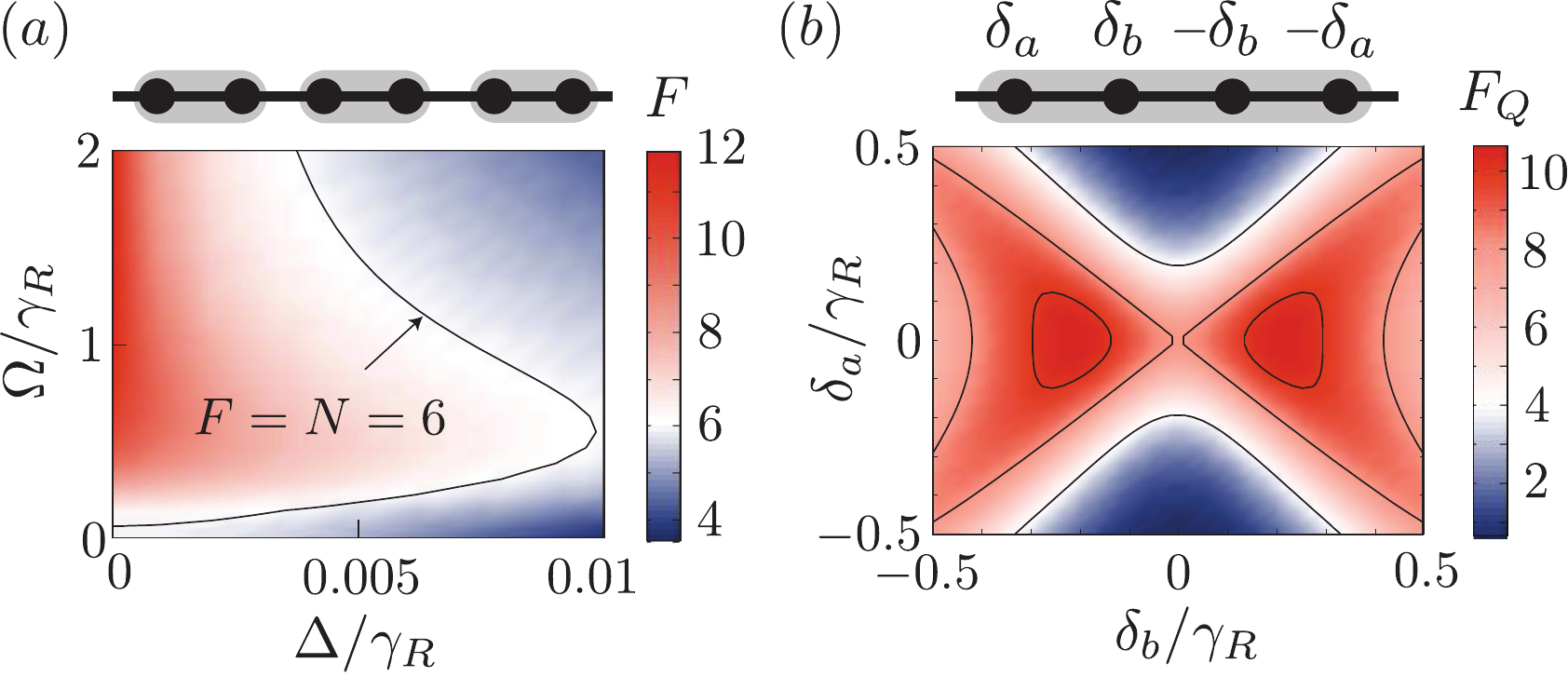}
\caption{(Color online) (a) Fisher information calculated via Eq.~(\ref{fisherdef}) for a generator $G=(1/2)\sum_j(-1)^j \sigma_j^{x}$ and a measurement of $J^z=\sum_{j} \sigma_{j}^z$, in the case of $N=6$ spins driven on resonance. The solid line shows the standard quantum limit $F=N$. (b) Quantum Fisher information calculated via Eq.~(\ref{quantumFisherInfo}), for a generator $G=(1/2)(\sigma_1^x-\sigma_2^x-\sigma_3^x+\sigma_4^x)$ in the case of $N=4$ spins driven with a strength $\Omega/\gamma_R=5$ and a detuning pattern $\delta_j=\{\delta_a,\delta_b,-\delta_b,-\delta_a\}$. The solid lines distinguish regions where $F_Q$  detects (at least) $n$-partite entanglement (for $n=2,3,4$). There is a parameter region where full four-partite entanglement is detected. All calculations are shown for a chirality of $\gamma_L=0.5\gamma_R$.}\label{fig:imperfections_fisher}
\end{figure}

\subsection{Quantum Fisher information for steady states of a chiral spin chain}

In this section we apply the above concepts to the different steady states of chiral spin networks studied in Sec.~\ref{Nspins}. In particular, we address the question of whether and up to what extent a measurement of the (quantum) Fisher information can reveal the multipartite entanglement structure of these states.

In the simplest example of a dimerized steady state, the entanglement stems from a finite overlap of each dimer with the singlet [cf.~Eq.~\eqref{Dimer}]. This singlet state is maximally sensitive to, e.g.,~staggered rotations around the $x$ axis [cf.~Fig.~\ref{fig:2spins_imperfection}(a)], and thus a suitable choice for the linear generator is given by $G=\frac{1}{2}\sum_j(-1)^j \sigma_j^{x}$. Moreover, it turns out that the measurement of the global operator $J^z=\sum_{j} \sigma_{j}^z$ is optimal to detect such rotations of a singlet state \cite{Wasak:2013ww}. 
A measurement of the corresponding Fisher information thus consists of two parts: (i) the generator $G$ is implemented by driving the spins on resonance with a coherent driving field, where in contrast to the homogeneous field used to drive the system into the dimerized state [cf.~Eq.~\eqref{Hsystem}], the amplitudes of this probe field have to be staggered, and (ii) determining the probabilities for the different measurement outcomes of the total spin along the $z$ axis, $J_z$. The Fisher information can then be calculated from this probability distribution for different values of the probe field (see Ref.~\cite{Strobel:2014eg}). 

In Fig.~\ref{fig:imperfections_fisher}(a) we calculate the corresponding Fisher 
information in the example of $N=6$ spins coupled by a chiral waveguide and driven on resonance. We additionally map out the region $F>N$, in which it detects bipartite entanglement. 
The Fisher information can witness entanglement even in the presence of imperfections such as a finite homogeneous detuning $\Delta$, and thus the steady state is neither pure nor dimerized. As one expects, no entanglement can be detected, if $\Delta$ is too large. In the ideal case ($\Delta=0$), and for strong driving $\Omega$, $F$ saturates at the maximum value $f(2,N)=2N$, consistent with a 2-producible state in which $N/2$ singlets are formed. 

\begin{figure}[h!!]
\includegraphics[width=0.46\textwidth]{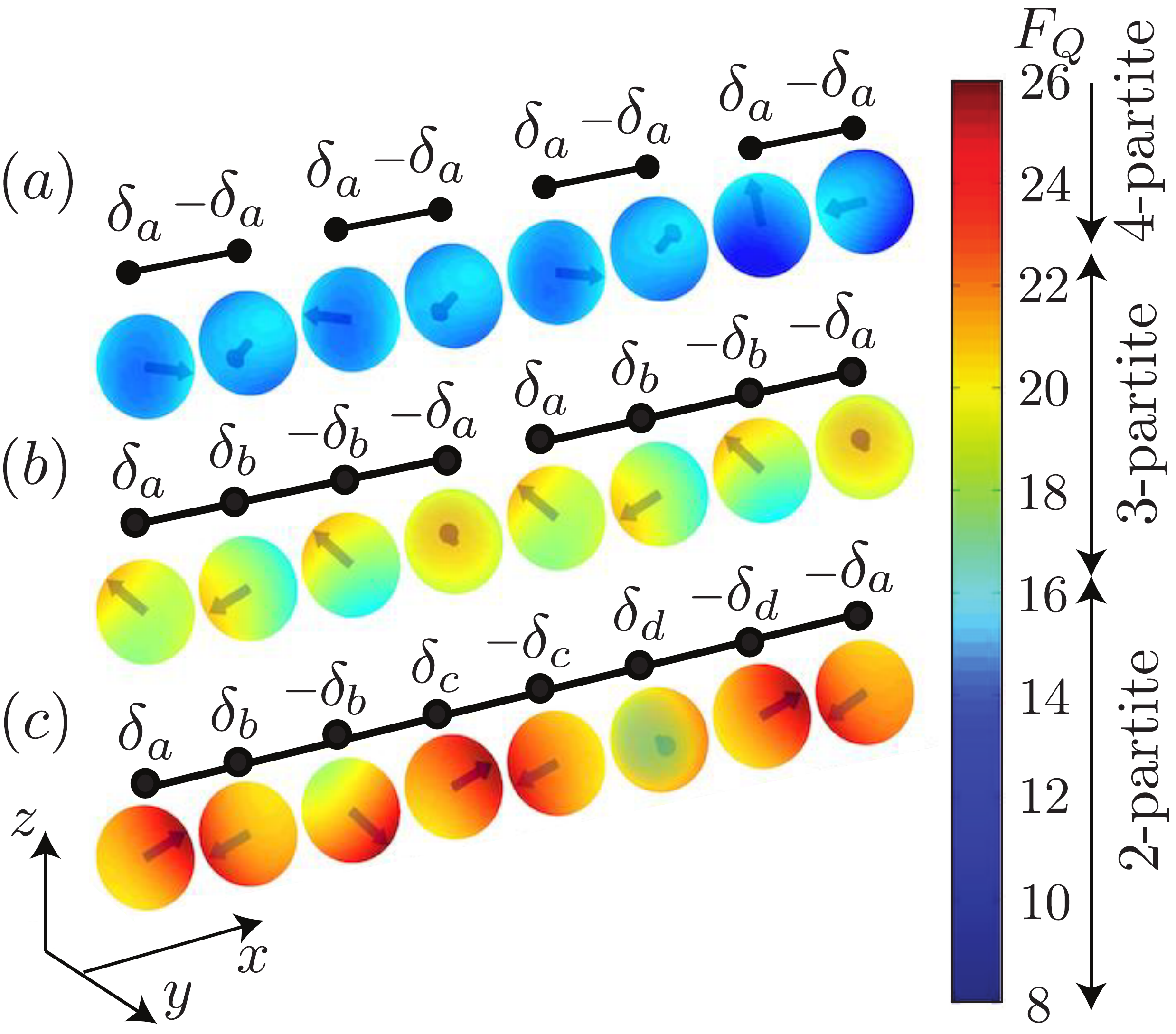}
\caption{(Color online) Optimal directions for local rotations to detect entanglement via the quantum Fisher information. In each panel (a)-(c) we indicate the directions $\vec{n}_i$ that give the maximum $F_Q$ for different steady states of a chiral spin network. We show examples of detuning patterns that give rise to (a) a dimerized state, (b) a tetramerized state, and (c) a fully eight-partite entangled octamer. The absolute values of the detunings are chosen (numerically) such that the steady state maximizes $F_Q^{\rm max}$. The color map on each sphere corresponds to $F_Q$ as a function of the local rotation direction $\vec{n}_i$, while keeping all other $\vec{n}_{j}$ ($j\neq i$) at their optimal value. One finds that $F_Q$ is able to detect the bipartite entanglement in the dimerized state, three-partite entanglement in the tetramerized state, and four-partite entanglement in the octamer. Other parameters are $\Omega/\gamma_R=5$ and $\gamma_L/\gamma_R=0.2$.}\label{fig:largechains}
\end{figure}

For dark states with a more complex entanglement structure than the dimerized state, such as the tetramer states for $N=4$ spins, it is not straightforward to analytically find generators and measurements that maximize the Fisher information. Nevertheless, the quantum Fisher information reveals that an optimal measurement can detect the full four-partite entanglement, for the simple generator $G=(\sigma_1^x-\sigma_2^x-\sigma_3^x+\sigma_4^x)/2$. This is shown in Fig.~\ref{fig:imperfections_fisher}(b) where $F_Q>f(3,4)=10$ in a specific parameter region. For tetramers corresponding to detunings outside this region, only three-partite, two-partite, or even no entanglement is witnessed. This reflects the fact that not every four-partite entangled state is equally useful for quantum metrology \cite{Hyllus:2010ix}.

To explore up to what extent the Fisher information can in principle detect the multipartite entanglement in the clustered states obtained in the chiral networks [cf.~Fig.~\ref{fig:largechains}], we employ \eref{Max_QFI} to numerically find the optimal generator and the corresponding maximal Fisher information for steady states of different structures. In Fig.~\ref{fig:largechains} we show this maximal $F_Q^{\rm max}$ for systems of $N=8$ spins, with detuning patterns leading to a dimerized state in Fig.~\ref{fig:largechains}(a), a tetramerized state in Fig.~\ref{fig:largechains}(b), and a full eight-partite entangled state in Fig.~\ref{fig:largechains}(c). The values of the detunings are thereby chosen such that the corresponding steady state maximizes $F_Q^{\rm max}$. In particular, we show the directions $\vec{n}_j$ for the optimal generator, obtained by a numerical solution of \eref{Max_QFI} and which give rise to the maximum quantum Fisher information. The color code on each sphere reflects the value of $F_Q$ as a function of the local generator direction $\vec{n}_i$ (where all other $\vec{n}_{j\neq i}$ are fixed at the optimal value), demonstrating the sensitivity of $F_Q$ against deviations from the optimal generator. In general, one finds that the optimal generator is not simple and in particular it is not a global one. An optimal measurement of the Fisher information therefore involves local rotations of the individual spins.  
As we have already discussed above, in the dimerized case, $F_Q$ easily detects the entanglement structure ($16>F_Q>8$) [Fig.~\ref{fig:largechains}(a)]. In higher entangled states, it detects the entanglement only partially [Figs.~\ref{fig:largechains}(b) and \ref{fig:largechains}(c)]. For example in the fully eight-partite state [Fig.~\ref{fig:largechains}(c)] up to four-partite entanglement is detected. 

\section{Summary and Outlook}\label{outlook}

In this paper we have discussed the driven-dissipative dynamics of a many-particle spin system interacting via a chiral coupling to a set of 1D waveguides. Our key result is the formation of pure, multipartite entangled states as steady states of the dynamics. Our results were derived within a quantum optical master equation treatment based on a Born-Markov elimination of the waveguides playing the role of quantum reservoirs. The emerging {\em many-body dark states} form clusters of spin states that decouple from the reservoir. The crucial ingredient of our scenario is the chirality of the reservoir, i.e.,~the symmetry breaking in the coupling of the spins to reservoir modes propagating in different directions. Moreover, we have shown that the multipartite entanglement emerging in these systems could be detected in a measurement of the Fisher information. 

We conclude with several remarks on the relevance and possible future extensions of the present work in a broader context. While the present work has focused on the non-equilibrium many-particle dynamics of chiral spin networks, we emphasize that chiral coupling of spins to waveguides has immediate applications in quantum communication protocols with spins representing the nodes (stationary qubits) of a network, connected by the exchange of photons (as flying qubits) \cite{Kimble:2008if}. In addition, we emphasize the Markovian assumption underlying our master equation treatment, which ignores time delays in the exchange of photons or phonons between spins. Inclusion of time delays and loops in these networks allows for a connection to quantum feedback problems \cite{WisemanBook}. Finally, it would be interesting to extend the present study to 2D geometries \cite{Neugebauer:2014iy,GonzalezTudela:2014wf}.

\section{Acknowledgments} We thank M. Oberthaler, B. Sanders, A. Smerzi and K. Stannigel for helpful discussions. Work in Innsbruck is supported by the ERC Synergy Grant UQUAM, the EU Grant SIQS, the Austrian Science Fund through SFB FOQUS, and the Institute for Quantum Information. T.~R. further acknowledges financial support from the BECAS CHILE scholarship program. Work at Strathclyde is supported by the EOARD through AFOSR Grant No. FA2386-14-1-5003.

\appendix
\
\section{Derivation of the Chiral Master equation}\label{DerivationME}

Here we derive the master equation for a collection of spins coupled to a chiral waveguide, as given in Eq.\,\eqref{chiralME}. We take a quantum optical point of view and identify the spins as the system and the bosonic modes in the chiral waveguide as the bath, which we will eliminate in a Born-Markov approximation \cite{QuantumNoise}. To do so it is convenient to  consider an interaction picture with respect to the bath Hamiltonian in Eq.\,(\ref{Hreservoir}), such that the total Hamiltonian in this frame reads $H_{\rm tot}(t)=H_{\rm sys}+H_{\rm int}(t)$. The density operator of the full system and bath at time $t$ is denoted by $W(t)$. Since the full system is closed, it simply evolves unitarily, $W(t)=U(t)W(0)U\dg(t)$. The unitary $U(t)$ satisfies the Schr\"odinger equation $i\hbar \frac{d}{dt}U(t)=H_{\rm tot}(t)U(t)$, with the initial condition $U(0)=1$. We choose the initial state of the system and bath as $W(0)=\rho(0)\otimes \ket{\rm vac}\bra{\rm vac}$; that is, system and bath are uncorrelated initially, and the bath is in the vacuum state. In the following we want to derive an equation of motion for the reduced density operator of the system $\rho(t)=\textrm{Tr}_B\{W(t)\}$, which is obtained from the state of the full system by tracing over the bath degrees of freedom. To this end we derive the quantum Langevin equations of motion, and from there we obtain the corresponding master equation. In a slightly more general situation than the one discussed in the main text, we allow here for the system bath couplings $\gamma_{\lambda}$ to vary from spin to spin, denoting the decay rate for spin $j$ by $\gamma_{\lambda j}$.

We start with the Heisenberg equations of motion for system operators $a(t)=U\dg(t)aU(t)$ and bath operators $b_{\lambda}(\omega,t)=U\dg(t) b_{\lambda}(\omega) U(t)$. The latter is given by
\begin{align}
\dot{b}_{\lambda}(\omega,t)=\sum_{l=1}^N \sqrt{\frac{\gamma_{\lambda l}}{2\pi}}\sigma_l(t)e^{-i(\nu-\omega)t-i\omega x_l/v_{\lambda}}
\end{align}
whose formal solution reads
\begin{align}
{b}_{\lambda}(\omega,t)={b}_{\lambda}(\omega)\!+\!\int_0^t \!\!ds  \sum_{l=1}^N \sqrt{\frac{\gamma_{\lambda l}}{2\pi}}\sigma_l(s)e^{i(\omega-\nu)s-i\omega \frac{x_l}{v_{\lambda}}},\label{solution_b}
\end{align}
with $b_{\lambda}(\omega,t=0)=b_{\lambda}(\omega)$.
The Heisenberg equations for an arbitrary operator $a$ acting on the Hilbert space of the spins only, read
\begin{widetext}
\begin{align}
\dot a(t)&\!=\!-\frac{i}{\hbar}[a(t),H_{\rm sys}(t)]
+\!\!\!\sum_{\lambda =L,R}\sum_{j=1}^N\int\!\! d\omega\sqrt{\frac{\gamma_{\lambda j}}{2\pi}}\!\lr{b_{\lambda}^{\dag}(\omega,t)e^{i(\omega-\nu)t -i\omega \frac{x_j}{v_{\lambda}}}[a(t),\sigma_j(t)]
-[a(t),\sigma_j\dg(t)]b_{\lambda}(\omega,t)e^{i(\nu-\omega)t +i\omega\frac{ x_j}{v_{\lambda}}}}.\label{Heisenberg_system}
\end{align}
Inserting the solution \eqref{solution_b} into Eq.~\eqref{Heisenberg_system}, denoting the quantum noise operators by $b_{\lambda}(t)\equiv\frac{1}{\sqrt{2 \pi}}\int d\omega\,b_{\lambda}(\omega)e^{-i(\omega-\nu)t}$, and introducing the shorthand notations $x_{jl}\equiv x_j-x_l$ and $k_{\lambda}\equiv \nu/v_{\lambda}$, one obtains
\begin{align}
&\dot a(t)=-\frac{i}{\hbar}[a(t),H_{\rm sys}(t)]+\sum_{\lambda =R,L}\sum_{j=1}^N\sqrt{\gamma_{\lambda j}}\lr{b_{\lambda}\dg(t-x_j/v_{\lambda})e^{-ik_{\lambda}x_j}
[a(t),\sigma_j(t)]-[a(t),\sigma_j\dg(t)]b_{\lambda}(t-x_j/v_{\lambda}))e^{ik_{\lambda} x_j}}\nonumber\\
&+\sum_{\lambda =R,L}\sum_{j,l=1}^N\frac{\sqrt{\gamma_{\lambda j}\gamma_{\lambda l}}}{2\pi}\int_0^t ds\int d\omega  \lr{e^{i(\omega-\nu)(t-s)-i\omega x_{jl}/v_{\lambda}} \sigma_l\dg(s)[a(t),\sigma_j(t)]-e^{-i(\omega-\nu) (t-s)+i\omega x_{jl}/v_{\lambda}}[a(t),\sigma_j\dg(t)]\sigma_l(s)}.\label{Heisenberg_system2}
\end{align}
\end{widetext}
\emph{Born-Markov approximation.} We assume that the timescales on which system operators evolve are much longer than the correlation time of the bath $\tau\sim1/\vartheta$. This is the essence of the Markov approximation \cite{QuantumNoise}, which allows us to perform the integrals over $\omega$ and $s$ in the second line of \eref{Heisenberg_system2}, assuming that $|\Omega_j|,|\delta_j|,\gamma_{\lambda j}\ll\vartheta\ll\nu $. For example (for times $t\!>\!|x_{jl}/v_{\lambda}|$), we obtain
\begin{align}
&\sum_{l}\int_0^t ds \int_{\nu-\vartheta}^{\nu+\vartheta} d\omega\, \frac{1}{2\pi} e^{i(\omega-\nu)(t-s)-i\omega x_{jl}/v_{\lambda}} \sigma_l\dg(s)\nonumber\\
&=\sum_l\int_0^t ds  \,\delta(t- x_{jl}/v_{\lambda}-s) e^{-ik_{\lambda}x_{jl}}\sigma_l\dg(s)\nonumber\\
&\approx\frac{1}{2}\sigma_l\dg(t)+\sum_l \theta(x_{jl}/v_{\lambda})e^{-ik_{\lambda}x_{jl}}\sigma_l\dg(t- x_{jl}/v_{\lambda}).
\end{align}
Here, the the function $\theta(x)$ is defined via $\theta(x)=1$ for $x>0$ and $\theta(x)=0$ for $x\leq0$ and it accounts for the time ordering of the spins along the two propagation directions.

\emph{Neglecting retardation.} In the following, we will further neglect retardation effects arising from a finite propagation velocity of the photons and approximate $\sigma_l(t- x_{jl}/v_{\lambda})\approx \sigma_l(t)$. This approximation is justified provided $| \Omega_j| ,| \delta_j|,\gamma_{\lambda j}\ll |v_{\lambda}|/|x_{jl}|$, that is, if time scales on which system operators evolve are much slower than the time photons need to propagate through the waveguide \cite{{Chang:2012co},Milonni:1974bo}. It is important to note that even though retardation effects are neglected, the time ordering of the spins along the waveguide is still accounted for. The ordering of the quantum noise operators $b_{\lambda }(\omega)$ allows a simple evaluation of expectation values $\langle a(t)\rangle={\rm Tr}_{\rm S+B}\{a(t)W(0)\}$ for initial states of the form $W(0)=\rho(0)\otimes\ket{\rm vac}\bra{\rm vac}$. Using the cyclic property of the trace and the fact that the bath is initially in the vacuum state ($b_{\lambda }(\omega)W(0)=W(0)b\dg_{\lambda }(\omega)=0$), one finds that the equation of motion for expectation values of arbitrary system operators $a$ is given by
\begin{widetext}
\begin{align}
\langle \dot a(t)\rangle&=-\frac{i}{\hbar}\langle [a(t),H_{\rm sys}(t)]\rangle+\sum_{\lambda =R,L}\sum_{j=1}^N\frac{\gamma_{\lambda j}}{2}\lr{\langle \sigma_j\dg(t)[a(t),\sigma_j(t)]\rangle -\langle [a(t),\sigma_j\dg(t)]\sigma_j(t)\rangle}\nonumber\\
&+\sum_{\lambda =R,L}\!\!\!\!\!\sum_{\substack{j,l\\k_{\lambda}x_j>k_{\lambda}x_l}}\!\!\!\!\!\sqrt{\gamma_{\lambda j}\gamma_{\lambda l}}\lr{e^{-ik_{\lambda}(x_j-x_l)}\langle \sigma_l\dg(t)[a(t),\sigma_j(t)]\rangle -e^{ik_{\lambda}(x_j-x_l)}\langle [a(t),\sigma_j\dg(t)]\sigma_l(t)\rangle}
\end{align}
We note that $\langle a(t)\rangle ={\rm Tr}_{\rm S+B}\{a(t)W(0)\}={\rm Tr}_{\rm S+B}\{aW(t)\}={\rm Tr}_{\rm S}\{a\rho(t)\}$; that is, one can move the time dependence in the expectation values for system operators to the reduced density operator. Since this above equation holds for all system operators, we obtain the master equation for the evolution of the system density operator $\rho(t)$ as 
\begin{align}
\dot \rho(t)&=-\frac{i}{\hbar}[H_{\rm sys},\rho(t)]+\sum_{\lambda =R,L}\sum_{j=1}^N\frac{\gamma_{\lambda j}}{2}\lr{[\sigma_j,\rho(t)\sigma_j\dg]-[\sigma_j\dg,\sigma_j\rho(t)]}\nonumber\\&+\sum_{\lambda =R,L}\!\!\!\!\!\sum_{\substack{j,l\\k_{\lambda}x_j>k_{\lambda}x_l}}\!\!\!\!\!\sqrt{\gamma_{\lambda j}\gamma_{\lambda l}}\lr{e^{ik_{\lambda}(x_j-x_l)}[\sigma_j,\rho(t)\sigma_l\dg]-e^{ik_{\lambda}(x_j-x_l)}[\sigma_j\dg,\sigma_l\rho(t)]}. 
\end{align}
\end{widetext}
Without loss of generality, we take $k_R\!=\!-k_L\!\equiv\! k\!>\!0$. This can always be achieved (for a single waveguide) by going to a different reference frame via the unitary transformation $V=\exp\lr{-i\frac{1}{2}\sum_{\lambda ,j}k_{\lambda} x_j \sigma_j^{\dag}\sigma_j}$. Simple algebra then shows that this is equivalent to the master equation presented in \eref{chiralME}. 

The bidirectional and the cascaded master equations in Eq.~(\ref{bidirectionalME}) and Eq.~(\ref{Cascaded_ME}) follow as special cases. It is further straightforward to generalize the above derivation to a chiral network and obtain the master equation \eqref{MultiME}. A chiral network as defined in Eq.~\ref{MultiME} simply consists of several independent chiral reservoirs, where the location of the spins along each of these waveguides may change. Since these reservoirs are all independent, the master equation is simply a sum of Liouvillians of the form in Eq.~\eqref{chiralME} for each waveguide. 

We note that we have performed a derivation of these equations by starting with a Hamiltonian in the rotating wave approximation. It is well known that the inclusion of the counterrotating terms is crucial to obtain the correct dipole-dipole interactions in three dimensions \cite{Lehmberg:1970jj}. One can derive the master equation also in 1D taking into account also the counterrotating terms \cite{GonzalezTudela:2013hn}; however, in 1D such a procedure leads to the same equation of motion \eqref{chiralME}.

\section{Explicit dark state for $N=4$ case}\label{tetramersolution}

In this appendix we give the explicit solution of the unique dark steady state of Eq.\,(\ref{Dickebroken}) in the case of $N=4$ spins discussed in Sec.~\ref{N4}. As commented on in Sec.\,\ref{N4}, the dark state can be found as an eigenstate of the coherent part of Eq.\,(\ref{Dickebroken}) within the subspace specified by \eref{explicit4state}, if the detunings fulfill at least one of the conditions (I)-(III) in Sec.~\ref{N4}.
Then, with $Q\!\equiv\!-i\Delta\gamma/2\!\neq\!0$, the five coefficients determining $\ket{\Psi}$ in Eq.\,(\ref{tetramer}) read
\begin{align}
a^{(1)}_{12}&\!=\!\frac{\Omega[2Q^2+2\delta_3\delta_4+(Q+\delta_1)(\delta_3+\delta_4)]}{\sqrt{2}(Q-\delta_1)(Q+\delta_3)(Q+\delta_4)},\label{alpha11}\\
a^{(1)}_{34}&\!=\!\frac{\Omega(2Q+\delta_3-\delta_4)}{\sqrt{2}(Q+\delta_3)(Q+\delta_4)},\\
a^{(1)}_{13}&\!=\!\frac{\Omega(\delta_3+\delta_4)}{2\sqrt{2}(Q+\delta_3)(Q+\delta_4)},\label{alpha13}\\
a^{(2)}_{1324}&\!=\!\frac{2\sqrt{2}\Omega{}{}a^{(1)}_{13}}{2Q-\delta_1-\delta_2},\label{alpha22}\\
a^{(2)}_{1234}&\!=\!\left\{\begin{array}{cc}\frac{\Omega^2(\delta_1+\delta_2-4Q)}{(Q-\delta_1)(Q+\delta_4)(\delta_1+\delta_2-2Q)}\!&,\ \textrm{ (I)  and (II)} \\ & \\ \frac{\sqrt{2}\Omega(4Q+\delta_3+\delta_4)a^{(1)}_{34}}{(2Q+\delta_3+\delta_4)(2Q-\delta_3+\delta_4)} &,\ {\rm (III)}\end{array}\right..\label{alpha21}
\end{align}
Notice that for the detuning pattern (I) the dark state factorizes into dimers $\ket{\Psi}\!=\!\ket{D(\alpha_1)}_{12}\ket{D(\alpha_3)}_{34}$, as defined in Eq.\,(\ref{Dimer}) with singlet fractions $\alpha_j\!\equiv\!-2\sqrt{2}\Omega/(2\delta_j+i\Delta\gamma)$. In the case of detuning patterns (II) and (III) the dark state is a genuine 4-partite entangled tetramer. On the other hand, when the bath is fully bidirectional ($\Delta\gamma=0$), dark state solutions of Eq.\,(\ref{Dickebroken}) also exist, provided all the detunings are nonzero and different (cf. Sec.\,\ref{nonlocaldimers}). The dark state is always dimerized and the specific detuning pattern determines how the spins pair up. For detunings (I), (II) and (III), it is given by $\ket{\Psi}\!=\!\ket{D(\alpha_1)}_{12}\ket{D(\alpha_3)}_{34}$, $\ket{\Psi}=\ket{D(\alpha_1)}_{13}\ket{D(\alpha_2)}_{24}$ and $\ket{\Psi}\!=\!\ket{D(\alpha_1)}_{14}\ket{D(\alpha_2)}_{23}$, respectively. Remarkably, in the last two cases the dimers are nonlocal [cf.~Sec.\,\ref{nonlocaldimers}].


\begin{thebibliography}{10}
\bibitem{footnote1} The two jump operators also coincide if $k|x_1-x_2|=(2n+1)\pi$ with $n=0,1,2,\dots$. Then the null space of $c_L=c_R=\sigma_1-\sigma_2$ is spanned by $\ket{gg}$ and $\ket{T}\!\equiv\!(\ket{ge}+\ket{eg})/\sqrt{2}$. This situation can be mapped to the one discussed in the text by the unitary transformation $V\!=\!\exp(-i\pi \sigma_2\dg \sigma_2)$.
\bibitem{footnote_cascaded} In general, the timescale $t_D$ in Eq.(\ref{Timescale_dimer}) for a dimer to be formed may vary from pair to pair due to possibly different detuning differences $\delta_{2j-1}\!-\!\delta_{2j}$, for $j=1,\dots,N/2$. The steady state is then reached on a timescale $t_{ss}\sim \sum_{j=1}^{N/2} t_{D}^j$.
\end{thebibliography}
\end{document}